\makeatletter
\declare@file@substitution{revtex4-1.cls}{revtex4-2.cls}
\makeatother

\documentclass[twocolumn,twocolappendix]{aastex631}

\usepackage{amsmath}
\usepackage{amssymb}
\usepackage{enumerate}

\newcommand{\be}{\begin{eqnarray}}
\newcommand{\ee}{\end{eqnarray}}
\newcommand{\rsun}{\ensuremath{R_{\odot}}}
\newcommand{\msun}{\ensuremath{M_{\odot}}}

\newcommand{\Eej}{\ensuremath{E_{\rm ej}}}
\newcommand{\Mej}{\ensuremath{M_{\rm ej}}}
\newcommand{\vmax}{\ensuremath{v_{\rm max}}}

\newcommand{\grad}{\ensuremath{\boldsymbol{\nabla}}}
\newcommand{\vel}{\ensuremath{\boldsymbol{v}}}
\newcommand{\rad}{\ensuremath{\boldsymbol{r}}}
\newcommand{\facev}{\ensuremath{\boldsymbol{w}}}
\newcommand{\qflux}{\ensuremath{\boldsymbol{Q}}}

\newcommand{\ddt}[1]{\ensuremath{\frac{\partial #1}{\partial t}}}

\newcommand{\flux}{\ensuremath{\boldsymbol{\mathcal{F}}}}

\newcommand{\state}{\ensuremath{\boldsymbol{\mathcal{U}}}}

\newcommand{\normal}{\ensuremath{\hat{\boldsymbol{n}}}}

\newcommand{\thermoD}[3]{\left(\frac{\partial #1}{\partial #2}\right)_{#3}}

\begin{document}

\title{Ejecta Wakes from Companion Interaction in Type Ia Supernova Remnants}


\correspondingauthor{Logan Prust}
\email{ljprust@kitp.ucsb.edu}

\author{Logan J. Prust}
\affiliation{Kavli Institute for Theoretical Physics, University of California, Santa Barbara, CA 93106, USA}

\author{Gabriel Kumar}
\affiliation{College of Creative Studies, University of California, Santa Barbara, CA 93106, USA}

\author{Lars Bildsten}
\affiliation{Kavli Institute for Theoretical Physics, University of California, Santa Barbara, CA 93106, USA}
\affiliation{Department of Physics, University of California, Santa Barbara, CA 93106, USA}

\begin{abstract}

Type Ia supernovae are triggered by accretion onto a white dwarf from a companion which is most likely Roche lobe-filling at the time of the explosion. The collision between the ejecta and a surviving companion carves out a conical wake, which could manifest as an asymmetry when the ejecta reaches the remnant phase. We simulate the companion interaction using the Athena++ hydrodynamics solver to determine the ejecta structure for a double-degenerate type Ia supernova. Ejecta in the wake is of lower density and higher velocity than the unperturbed ejecta. We then evolve the ejecta for several thousand years using the expanding-grid code Sprout. The forward shock within the wake is initially indented, but becomes spherical after roughly a thousand years due to transverse motion of shocked ejecta that fills the wake. The reverse shock travels quickly within the wake, leading to an off-center convergence of the reverse shock and leaving the remnant with an asymmetrical core. This also draws material from the interstellar medium deep into the remnant, eventually reaching the center. Large Rayleigh-Taylor plumes are found around the edge of the wake, creating a toroidal structure composed primarily of ejecta. Estimates of the thermal X-ray emission show that such remnants exhibit observable asymmetries for thousands of years.

\end{abstract}


\section{Introduction} \label{sec:intro}

Type Ia supernovae (SNIa) are thought to be triggered by either the merger of two white dwarfs or by accretion onto a white dwarf (WD) from a binary companion. For a review of explosion mechanisms, see \citet{2023RAA....23h2001L}. In the latter case, there are a variety of possibilities for the nature of the binary donor, but because the donor is Roche-lobe filling the solid angle it subtends relative to the accretor is nearly identical. This guarantees that a significant portion of the ejecta will collide with the donor, which may have a lasting effect on both the donor and ejecta.

Several works have explored the response of the donor to mass stripping and entropy deposition from the traversed shock \citep{2000ApJS..128..615M,2015A&A...584A..11L,2015MNRAS.449..942P,2018ApJ...864..119H,2018ApJ...868...90T,2019ApJ...887...68B,2024arXiv240800125W} in both single-degenerate (SD) and double-degenerate (DD) binaries. In the DD scenario, some authors found that the donor detonates \citep{2015MNRAS.449..942P,2019ApJ...885..103T}, while others found a shocked donor consistent with the hypervelocity white dwarfs observed by Gaia \citep{2018ApJ...865...15S,2023OJAp....6E..28E}.

The collision also forms a bow shock which carves out a wake in the ejecta filled with low-density, shock-heated gas \citep{2010ApJ...708.1025K}. We explore here to what degree this asymmetry carries over to the remnant phase, when the mass of interstellar medium (ISM) swept up by the forward shock is comparable to the ejecta mass. SNIa remnants have been found to display more spherical symmetry and mirror symmetry than core-collapse remnants \citep{2011ApJ...732..114L}, though significant asymmetries have been observed in Ia remnants such as Tycho \citep{2023A&A...680A..80G} and Kepler \citep{2024A&A...687A..28P}. \citet{2024arXiv240312264M} found that the power spectrum of Tycho has a low-wavenumber peak which could be explained by companion interaction, though they note that a large-scale density gradient in the circumstellar medium or a velocity kick during the SN could also be responsible. Many of these objects are targeted for observations by the X-Ray Imaging and Spectroscopy Mission \citep{2024arXiv240814301X} in the coming years.

Supernova remnants (SNRs) resulting from SNIa ejecta in the SD case were investigated by the axisymmetric smoothed-particle hydrodynamics simulations of \citet{2012ApJ...745...75G} and \citet{2016ApJ...833...62G}, focusing on the first few centuries of evolution. \citet{2012ApJ...745...75G} found large Rayleigh-Taylor (RT) plumes at the edge of the wake, nearly reaching the forward shock. \citet{2016ApJ...833...62G} also found that the remnant was clearly asymmetrical in X-rays at both $t=100$ and 300 years, exhibiting a ring around the edge of the wake that is visible from any viewing angle. \citet{2022ApJ...930...92F} tackled the DD case using the Eulerian code RAMSES \citep{ramses}, finding that the ejecta tail from the helium shell detonation produces a protrusion at early times but the effects of the ejecta wake lasted much longer.

Though the studies above have identified important features of the shock surfaces and X-ray emission, it is unclear how the late-time ($\gtrsim$1000 yrs) evolution of the shocks affect the SNR morphology and composition, particularly the self-interaction of the reverse shock as it reaches the center of the remnant at $t\sim 1000$ yrs. To this end, we perform hydrodynamical simulations of the collision between the SNIa ejecta and companion to obtain the wake structure. We then homologously evolve the resulting ejecta structure to the remnant phase and compute its subsequent 3-D evolution using a Cartesian expanding-grid code.

We organize this paper as follows. The setup for our ejecta collision simulation is described in section \ref{sec:athenasetup}, and the results are presented in section \ref{sec:athenaresults}. The setup for evolving the subsequent ejecta structure through the remnant phase are shown in section \ref{sec:sproutsetup}. The 3-D dynamics of the resulting remnant evolution are presented in section \ref{sec:shocks} and compared to semi-analytic predictions. Synthetic observations of thermal X-ray emission are shown in section \ref{sec:tomography}, and in section \ref{sec:discussion} we discuss our results and conclude.


\section{Numerical Setup for Companion Interaction} \label{sec:athenasetup}

We perform hydrodynamical simulations using Athena++ \citep{athena++}, which solves the Euler equations
\be
\ddt{\rho} + \grad\cdot\rho\vel &=& 0, \label{eq:continuity}\\
\ddt{\rho\vel} + \grad\cdot\rho\vel\vel + \grad P &=& 0, \label{eq:momentum}\\
\ddt{u} + \grad\cdot\left(e + P\right)\vel &=& 0 \label{eq:energy}
\ee
for material with density $\rho$, fluid velocity $\vel$, pressure $P$, energy density $e=\epsilon+\rho v^{2}/2$, and internal energy $\epsilon$. Fluxes are computed using a Harten–Lax–van Leer contact approximate Riemann solver \citep{1994ShWav...4...25T}. A piecewise linear method (PLM) with a van Leer slope limiter is used for spatial reconstruction with the corrections described in \citet{2014JCoPh.270..784M} to keep the limiter total variation diminishing. Time integration is handled via a second-order van Leer predictor-corrector scheme with the Courant–Friedrichs–Lewy number set to $\eta_{\rm CFL}=0.3$.

The SNIa ejecta is gas-pressure dominated at a few seconds after the detonation (Kumar et al., in preparation), but may become radiation-pressure dominated when shocked. Thus, we treat the gas as ideal and include radiation pressure assuming local thermodynamic equilibrium. Then the internal energy and pressure are given by
\be
\epsilon &=& a_rT^4+(\gamma-1)\frac{\rho k_B T}{\mu m_p} \label{eq:energyDensity},\\
P &=& \frac{1}{3}a_rT^4+\frac{\rho k_B T}{\mu m_p} \label{eq:pressure},
\ee
where $T$ is the gas and radiation temperature, $a_r$ is the radiation constant, $\gamma$ is the specific heat ratio, $k_B$ is the Boltzmann constant, $m_p$ is the proton mass, and we choose $\mu=4/3$ for the mean molecular weight of the fully-ionized plasma. Closure of the Euler equations requires conversion between pressure and energy, and thus a determination of $T$. We accomplish this via a quartic solver which performs a root find to solve equations of the form $T^4+BT-A=0$. We also require a determination of the adiabatic sound speed $c_s^2=\gamma P/\rho$. Here $\gamma$ is given by
\be
\gamma=\thermoD{P}{\rho}{S}=\frac{32-24\beta-3\beta^2}{24-21\beta} \label{eq:gamma}\\
\ee
\citep{1984oup..book.....M}, where $\beta=P_{\rm gas}/P$ and $P_{\rm gas}=\rho k_B T/\mu m_p$.



We use a 3-D spherical-polar mesh with the explosion at the origin. The domain extends from an inner radial boundary $R_{\rm in}=0.129\rsun$ -- where we inject the SN ejecta -- to an outer boundary at $R_{\rm out}=10.7\rsun$ where the gas freely outflows. This $R_{\rm out}$ is sufficiently large to ensure that the gas exiting there has reached homology, which we demonstrate in section \ref{sec:athenaresults}. The donor is located on the $\theta=0$ pole at an orbital separation of $a=0.269\rsun$ with radius $R=0.0795\rsun$, which is similar to model 1 of \citet{2019ApJ...887...68B} and model HeStar1 of \citet{2024arXiv240800125W}. The domain extends out to a maximum polar angle of 1.4 radians, while the azimuthal angle $\phi$ spans the full range of $0$ to $2\pi$. Our mesh spacing is uniform in the polar and azimuthal directions, whereas in the radial direction we use geometric spacing with neighboring cells differing in size by a factor of 1.0013. In total, our domain contains $N_{r}\times N_{\theta}\times N_{\phi} = 2160\times 200\times 10 = 4320000$ cells. Outside of the ejecta is an ambient medium with density $\rho_0=10^{-7}$ $\textrm{g/cm}^3$ and pressure $P_0=5\times 10^8$ g/(cm s$^{2}$). The total mass of this medium within the domain is $\approx 2\times 10^{-5}\msun$ which is lower than the mass of the ejecta inserted into the domain by four orders of magnitude.

Though all SNIa are expected to have a Roche lobe-filling companion, we also investigate cases in which the orbital separation is increased by a factor of $\eta$ (without changing the donor size) to explore the limiting case. Specifically, we perform simulations with $\eta=\sqrt{3}$ and $\eta=3$, corresponding to reductions in the solid angle subtended by the donor of $\eta^{2}=3$ and 9, respectively. We also scale the radial size of the mesh and all time scales by $\eta$ and reduce the pressure and density of the ambient medium by $\eta^{3}$, preserving its total mass and sound speed.

\subsection{Injection of SN Ejecta} \label{sec:BCs}

The supernova ejecta is injected into our domain at the inner radial boundary $R_{\rm in}$. The radial profile of the ejecta is adopted from the Gaussian fit of \citet{2024arXiv240800125W}: 
\be
\rho_{G}(v,t)=\frac{\Mej}{(v_{0}t\sqrt{\pi})^{3}} \exp(-v^{2}/v_{0}^{2}), \label{eq:densityProfileVelocity}
\ee
where $t$ is the time since shock breakout, $\Mej$ is the ejecta mass, and $\Eej$ is the total energy of the ejecta. The characteristic velocity $v_{0}$ is given by
\be
v_0=\sqrt{\frac{4\Eej}{3\Mej}}. \label{eq:v0}
\ee
\citet{2024arXiv240800125W} showed that this form is a better representation of the ejecta's velocity structure than previous choices such as broken power laws or exponentials. We choose an ejecta mass of $\Mej=0.9\msun$ and an explosion energy of $\Eej=0.97\times10^{51}$ ergs based on the explosion model of \citet{2018ApJ...854...52S} for a WD with a 50/50 C/O ratio. As the majority of the kinetic energy is carried by gas with $v\approx v_{0}$, we truncate the ejecta at a maximum velocity of $\vmax=3v_{0}$. The pressure is set assuming initially gas pressure-dominated ejecta with uniform entropy: $P=K \rho^{5/3}$. The pseudoentropy $K=7\times10^{13}$ cm$^{4}$/g$^{2/3}$s$^{2}$ is chosen to be within the characteristic range of the detonation simulations. As the velocity of the injected gas decreases, the stand-off distance of the bow shock from the donor increases, eventually reaching $R_{\rm in}$ at $t\approx 100$ s. This is two orders of magnitude larger than the timescale $R_{\rm in}/v_{0}$, meaning that the gas entering the domain after this time is far behind the outer layers of the ejecta which are the focus of this paper.



\subsection{Treatment of the Donor}

We treat the donor as a reflective spherical boundary. This neglects the small transverse motion of the donor relative to the ejecta as well as internal shocks (and subsequent ringing) within the donor. However, the sound-crossing time of the donor (tens of seconds) is long enough that the bulk of the outermost ejecta has passed by the donor prior to shock breakout from the back of the donor \citep[see][]{2024arXiv240800125W}. Thus, we do not expect large changes to the morphology of the ejecta from mass loss due to internal shocks. Modeling the donor as a perfect sphere also neglects tidal deformation prior to the SN, though \citet{2024arXiv240800125W} find that in practice this produces a negligible effect on their results. We also neglect material stripped from the donor, though this is insignificant in the DD case \citep{2024arXiv240800125W} and has been shown to be insufficient to fill in the wake in the SD case by \citet{2012ApJ...745...75G} and \citet{2016ApJ...833...62G}.

To implement this reflective boundary, we modify the Riemann solver such that the cells at the surface of the boundary are treated as having fluid states which are locally equal to that of the adjacent gas, but with opposite fluid velocity normal to the surface. With PLM interpolation, this ensures that all numerical fluxes vanish at the surface with the exception of the momentum flux normal to the surface. The fork of Athena++ used to perform these calculations can be found at \url{https://github.com/ljprust/athena/tree/sn1a}, using the problem generator src/pgen/supernovaCollision.cpp and parameter file inputs/hydro/athinput.supernovaCollision.

\section{Companion Interaction Results} \label{sec:athenaresults}

\begin{figure}
    \centering
    \includegraphics[width=1.0\linewidth]{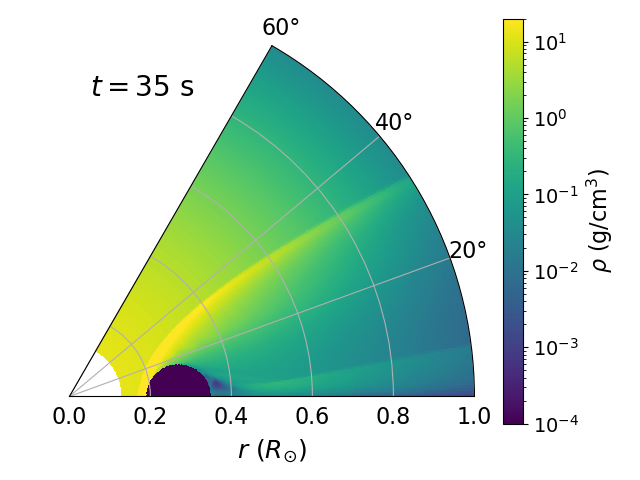}
    \includegraphics[width=1.0\linewidth]{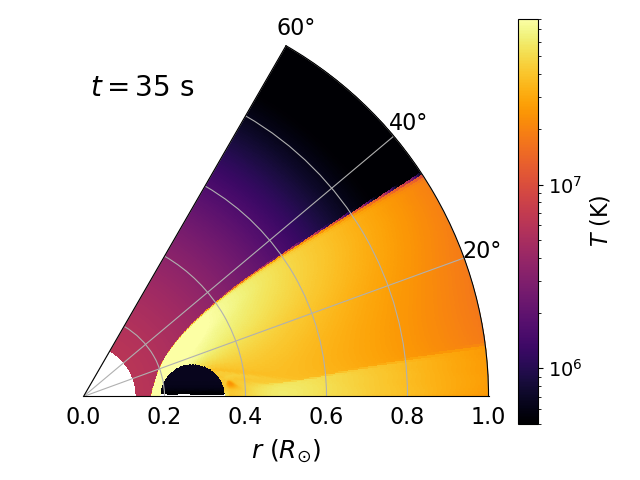}
    \caption{Slices of density (top) and temperature (bottom) at the $\phi=0$ plane in our fiducial Athena++ run, showing the bow shock and recompression shock.}
    \label{fig:athenaslice}
\end{figure}

\begin{figure}
    \centering
    \includegraphics[width=1.0\linewidth]{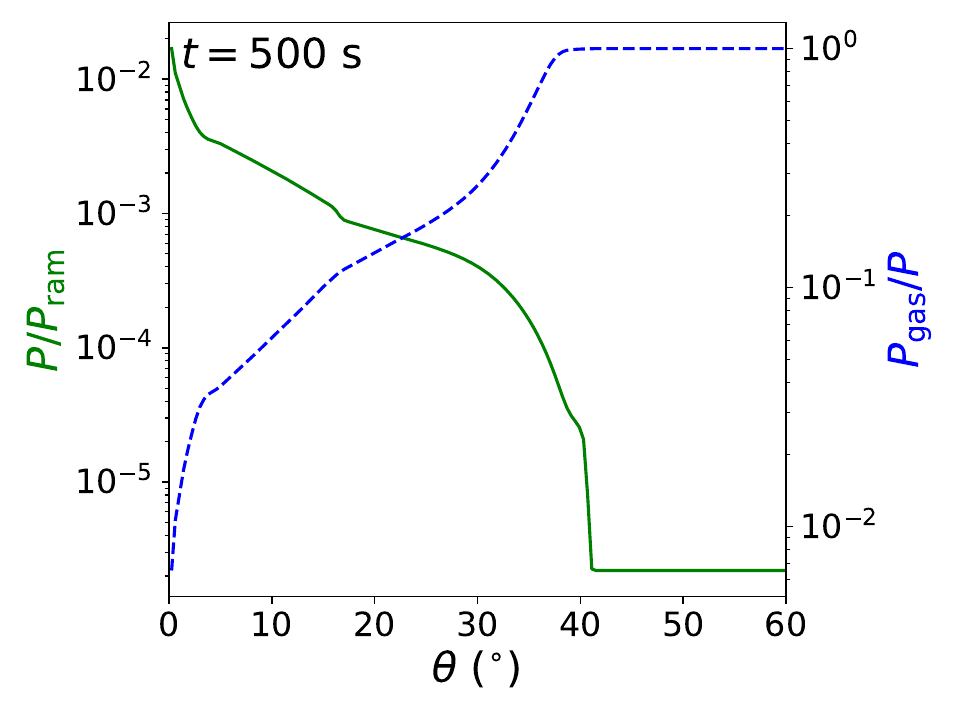}
    \caption{Ratio of total pressure (gas + radiation) to ram pressure (solid green line, left axis) and gas pressure to total pressure (dashed blue line, right axis) at the outer radial boundary of the domain at $t=500$ s. The total pressure is negligible compared to the ram pressure for all $\theta$, confirming that the gas is in homology. The gas outside of the wake remains firmly gas-pressure dominated, while much of the wake is radiation-pressure dominated.}
    \label{fig:Pratio}
\end{figure}

\begin{figure}
    \centering
    \includegraphics[width=1.0\linewidth]{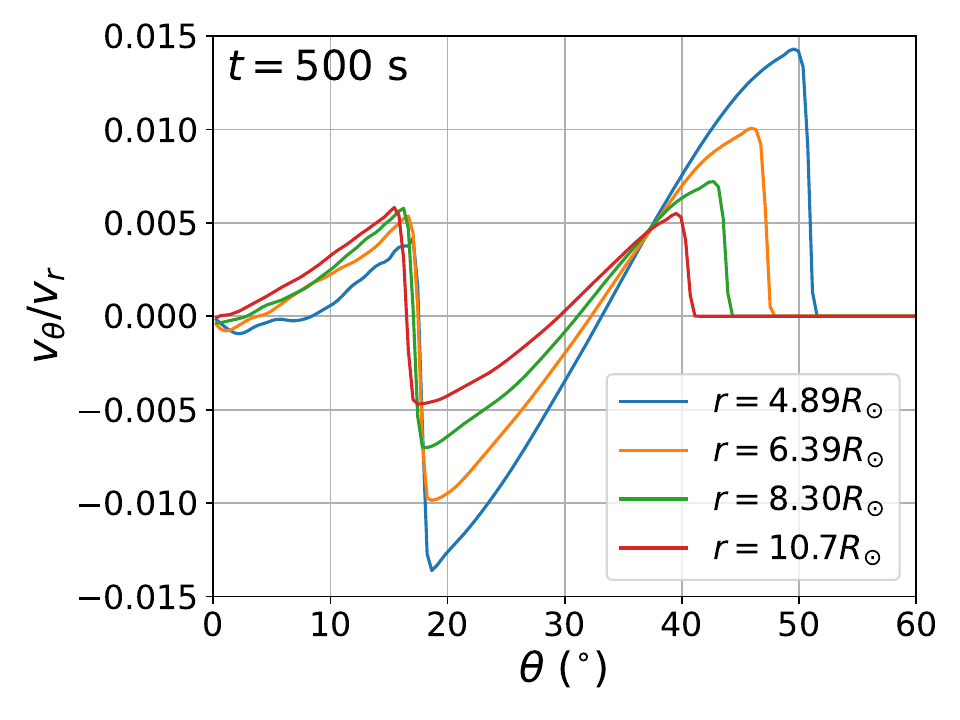}
    \caption{Ratio of polar to radial velocities at $t=500$ s, showing discontinuities at both shocks with a rarefaction wave in between. For large $r$, $v_{\theta}/v_{r}<0.01$ is satisfied.}
    \label{fig:vtheta}
\end{figure}

The setup described above is used to simulate the collision between the ejecta and companion until $t=1000$ s. Slice plots of the density and temperature are shown in Fig.~\ref{fig:athenaslice}, revealing the presence of a recompression shock downstream of the donor. The temperature of the shocked ejecta is 1 to 3 orders of magnitude larger than that of the unshocked ejecta at any given radius. Because of this, the shocked ejecta is almost entirely radiation-pressure dominated, as demonstrated in Fig.~\ref{fig:Pratio} (dashed blue line) at $R_{\rm out}$ at a later time of $t=500$ s. The solid green line shows the ratio of total pressure (gas + radiation) to ram pressure which is low for all of the ejecta, confirming that it is in homology as it exits the domain. 

The fluid velocity in the polar direction (Fig.~\ref{fig:vtheta}) also shows the bow and recompression shocks as well as a rarefaction wave between them and demonstrates that the flow can be approximated as radial at $R_{\rm out}$. This figure also elucidates the widening of the bow shock with time, reaching $\theta>50^{\circ}$. This means that at least $\approx$18\% of the ejecta (by solid angle) has experienced a degree of shock heating.

\subsection{Extrapolation to the Remnant Phase} \label{sec:extrapolation}

\begin{figure}
    \centering
    \includegraphics[width=1.0\linewidth]{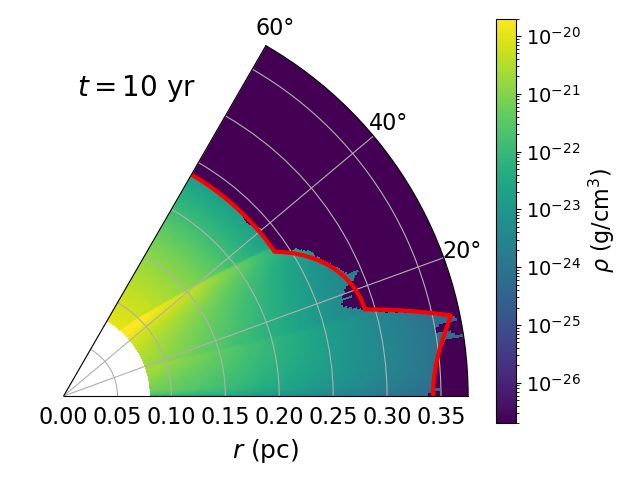}
    \caption{Homologously-evolved ejecta and our fit for the ejecta surface (red line). The velocity of the unperturbed ejecta is exceeded by the ejecta which passed through the bow shock, and is higher still for ejecta which passed through both shocks.}
    \label{fig:homologous}
\end{figure}

We homologously evolve the ejecta from $t=1000$ s to $t=10$ yr, at which time the mass of the ISM swept up by the forward shock starts to matter given $n_{\rm ISM}\sim 1$ proton per cm$^{3}$. Because the ejecta is homologous, each parcel of gas can be treated as a Lagrangian particle whose density evolves as $\rho\propto t^{-3}$. The treatment of internal energy is far more complicated due to radioactive heating from $^{56}$Ni, though for our purposes this is irrelevant as the internal energy is effectively zero by the start of the remnant phase. The results of this extrapolation are shown in Fig.~\ref{fig:homologous}. Here we see that the shocked ejecta has been accelerated by its interaction with the donor, and thus the outer surface of the ejecta is not spherical. The gas which passed through both shocks experiences the greatest acceleration, up to $\approx$50\% of its initial velocity. We approximate the outer surface of the ejecta with a fitting function (red line), the details of which are given in Appendix \ref{sec:fits}.

\subsection{Single-Degenerate Model} \label{sec:kasen}

\begin{figure}
    \centering
    \includegraphics[width=1.0\linewidth]{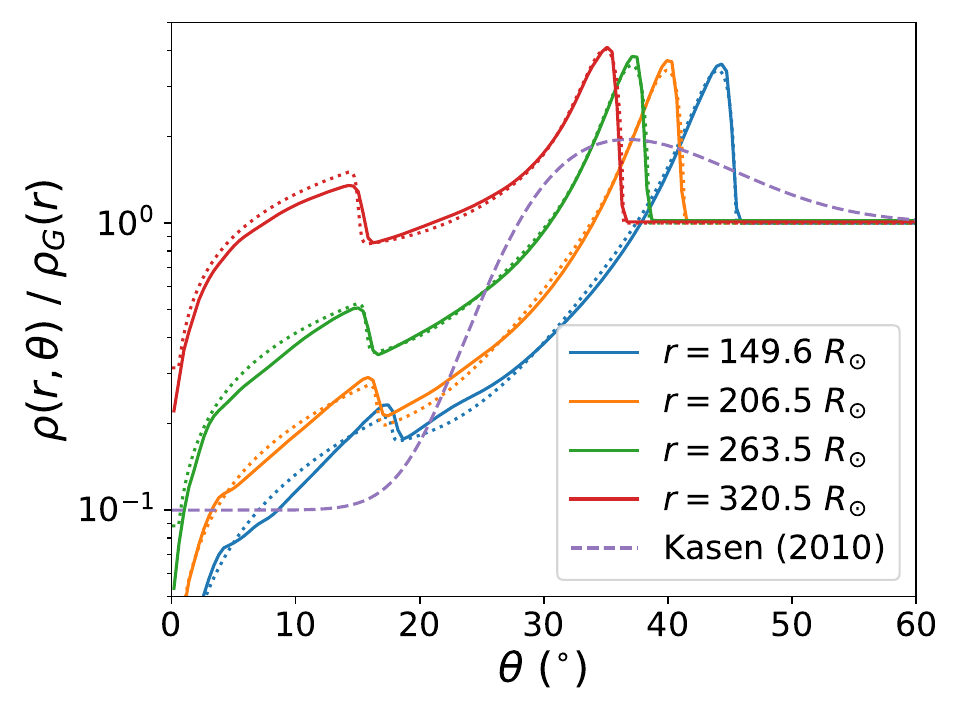}
    \caption{Comparison between ejecta profiles in our fiducial Athena++ run (solid lines) with our fitting functions (dotted lines) at $t=10000$ s. Each line is normalized by the density of unshocked ejecta at that radius. This is contrasted to the smoother \citet{2010ApJ...708.1025K} SD model (dashed line) which does not contain a recompression shock.}
    \label{fig:kasenvsathena}
\end{figure}

As a comparison to our double-detonation model, \citet{2010ApJ...708.1025K} simulated the interaction between SNIa ejecta and a red giant companion. They obtained a fit to their ejecta structure which separates the radial and azimuthal profiles:
\be
\rho(r,\theta) = \rho(r) f_{0}(\theta),
\ee
where
\be
\begin{aligned}
f_{0}(\theta) &= f_{h}+(1-f_{h})\frac{x^{m}}{1+x^{m}} \\ &\times\left(1+A\exp\left[-\frac{(x-1)^{2}}{(\theta_{p}/\theta_{h})^{2}}\right]\right). \label{eq:kasen}
\end{aligned}
\ee
Here $x=\theta/\theta_{h}$ and the fit parameters are $m=8$, $f_{h}=0.1$, $\theta_{h}=30^{\circ}$, $\theta_{p}=15^{\circ}$, and $A=1.8$. This is plotted against our data in Fig.~\ref{fig:kasenvsathena}, showing several qualitative differences such as the lack of a recompression shock and the uniformity of the density at small $\theta$. Here we also plot fits to the Athena++ data (dotted lines), the details of which are given in Appendix \ref{sec:fits}. These fits allow us to extrapolate the Athena++ data to $r=0$, though this results in a total remnant mass of 0.892$\msun$ which is slightly lower than our chosen ejecta mass of 0.9$\msun$. The SD result (\ref{eq:kasen}) combined with the set of fitting functions describing our DD results provide two ejecta structures which we evolve through the remnant phase.

\section{Numerical Setup for Remnant Evolution} \label{sec:sproutsetup}

The SNR evolution is computed using the 3-D moving-mesh code Sprout \citep{sprout}, which has already been applied to several studies of SNRs \citep{2023ApJ...956..130M,2024arXiv240312264M}. Sprout uses a uniformly-expanding mesh, which allows the evolution of a SNR to be followed over multiple orders of magnitude without remapping onto a new grid. We set the expansion speed of the mesh such that the largest fluid velocity present locally matches the mesh speed, which typically occurs at either the forward or reverse shock. However, as we will show in section \ref{sec:shocks}, the reverse shock can drive large fluid velocities near the center of the remnant and particularly in the wake. This dramatically accelerates the mesh speed, leading to numerical errors at the forward shock. For this reason, when computing the mesh speed we ignore the fluid in the wake and near the center of the remnant. A Runge-Kutta second-order scheme is used for time integration.

We use an HLLC Riemann solver to compute the numerical fluxes. As shown in \citet{sprout}, on a uniformly-expanding grid the change in cell volumes and face areas can be included directly into the flux computation: 
\be
\flux_{m}=C_{F}\flux_{s}-C_{U}\state\facev^{T}. \label{eq:fluxm2}
\ee
Here $\flux_{s}$ is the flux computed as if the mesh was static, $\state$ is the fluid state, and $\facev$ is the local mesh velocity. The dyad $\state\facev^{T}$ represents the gas ``swept up'' by the motion of a cell face. The coefficients $C_{U}$ and $C_{F}$ account for changes in cell volume and face area, which can be computed analytically when the mesh expansion is uniform. This differs from unstructured grids, where areas and volumes must be measured at each time step and subsequently applied to the flux. One issue faced by moving-mesh codes is that the two terms on the right-hand side of equation (\ref{eq:fluxm2}) nearly cancel out since the flux velocity and mesh velocity are often similar, leading to sign errors in $\flux_{m}$. We modify the Riemann solver such that the flux is computed in the local rest frame of each cell interface, which reduces such errors. The formalism for this transformation is described in Appendix \ref{sec:upwind}, though we find only modest morphological differences in the SNR between the two methods.

Sprout also includes a low-Mach (LM) HLLC solver \citep{2020JCoPh.42309762F}, which reduces numerical diffusion in the presence of low directional Mach number in order to eliminate carbuncle instabilities in grid-aligned shocks. We find that material in the wake can erroneously trigger the LM correction, and thus we opt to use the HLLC solver without LM modifications. This results in small carbuncles forming at points at which the forward shock is aligned with the grid, though these appear to have a negligible impact on the results. The fork of Sprout used for these calculations can be found at \url{https://github.com/ljprust/Sprout/tree/upwind}.

In contrast to our Athena++ runs, here we model a quadrant of the ejecta; that is, $\theta\in [0,\pi]$, $\phi\in [0,\pi/2]$. Our Cartesian grid contains $N_{x}\times N_{y}\times N_{z}=512\times 512\times 1024=268$ million cells, with the $\theta=0$ pole aligned with the $+z$-axis. The initial size of our domain is chosen such that $L_{x}=L_{y}=L_{z}/2$ is roughly twice the initial radius of the unperturbed ejecta. Hereafter we will use ``unperturbed'' to refer to the portion of the ejecta which has been altered by the companion interaction rather than ``unshocked'' to avoid confusion regarding material shocked by the forward and reverse shocks. The center of the remnant lies at $[0,0,L_{z}/2]$, which is also the center point for the mesh expansion.

The gas pressure is initialized as $P=10^{-5}\rho v_{\rm max}^{2}$ so that ram pressure dominates over gas pressure and $v_{\rm max}\gg c_{s}$. We ignore radiation pressure and treat the gas as ideal with $\gamma=5/3$. Outside of the ejecta, we fill the domain with an ISM with density $6.31\times 10^{-25}$ g/cm$^{3}$, which is chosen to roughly match the ISM density for the SNIa remnant SNR 0509-67.5 as inferred by \citet{2022ApJ...938..121A}. We include a passive tracer to track the mass fraction of ejecta. \\\\



\section{Dynamical Remnant Evolution} \label{sec:shocks}

\begin{figure*}
\begin{tabular}{llll}
\includegraphics[height=1.1\linewidth,trim={2cm 0 2cm 0},clip]{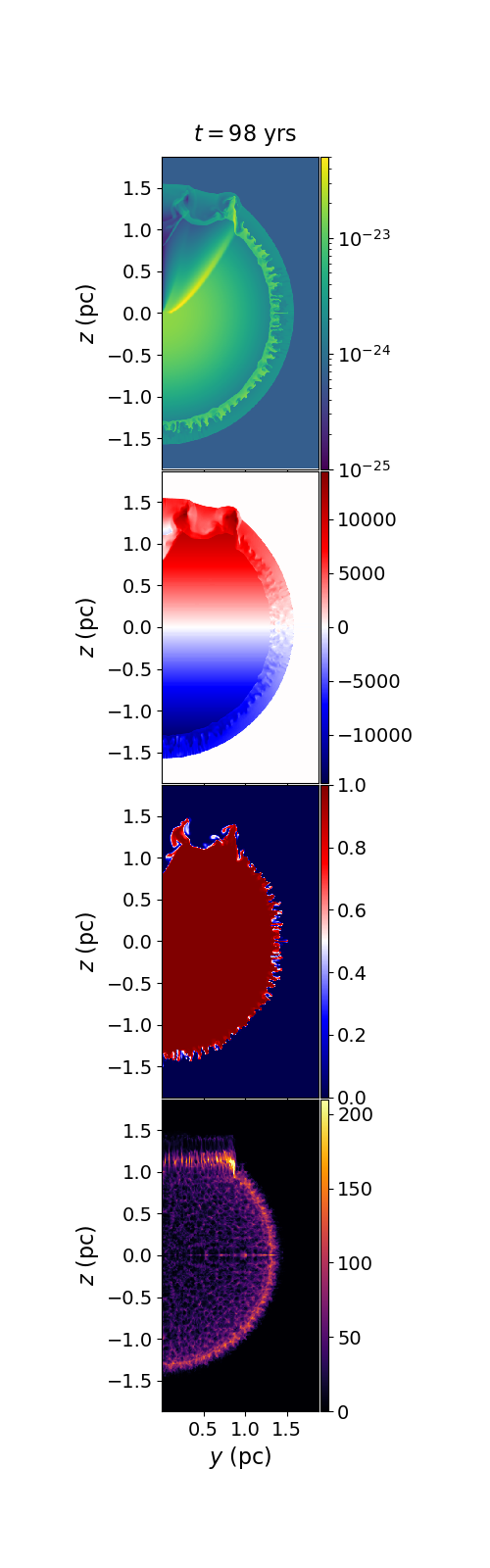} & 
\includegraphics[height=1.1\linewidth,trim={3cm 0 2cm 0},clip]{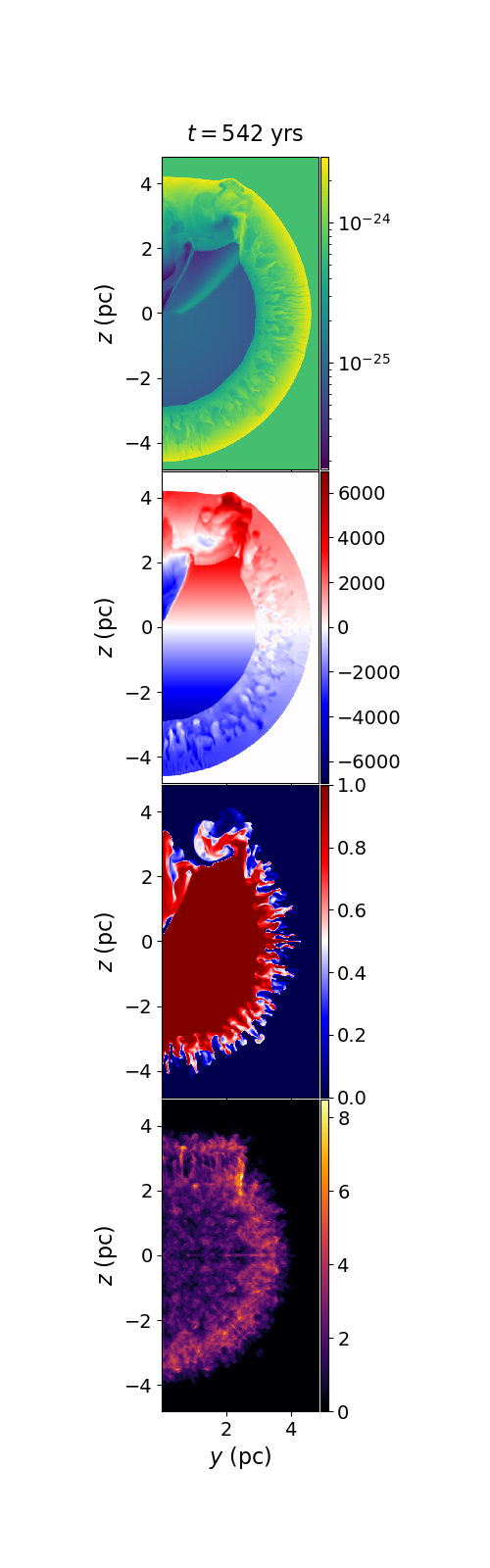} &
\includegraphics[height=1.1\linewidth,trim={3cm 0 2cm 0},clip]{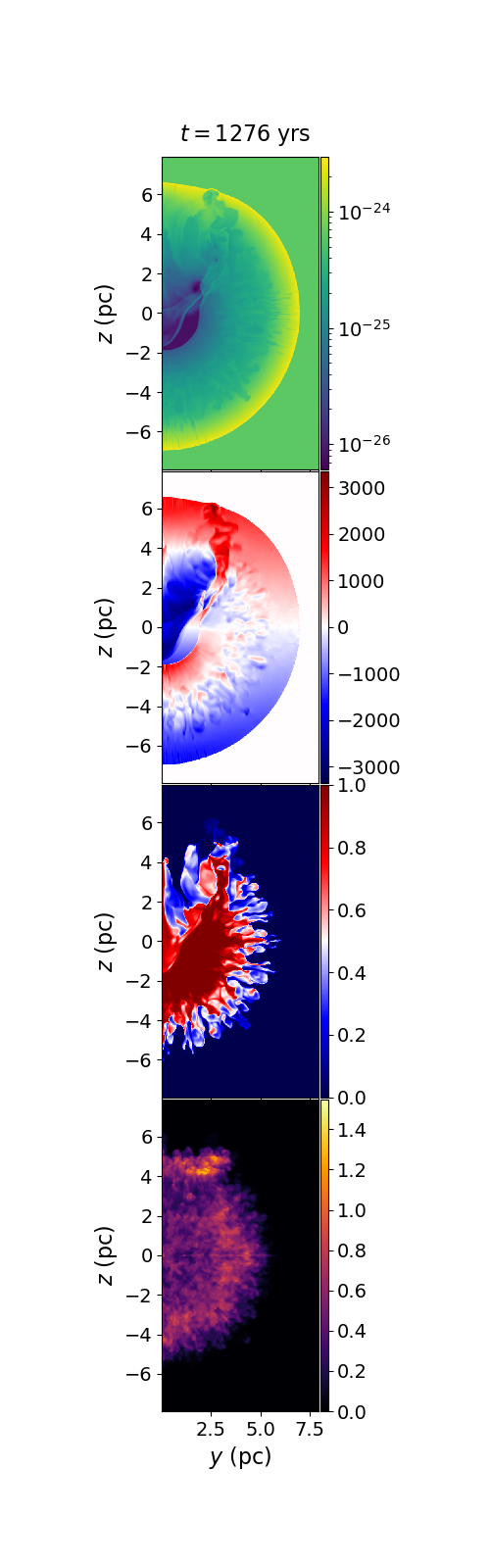} &
\includegraphics[height=1.1\linewidth,trim={3cm 0 1.5cm 0},clip]{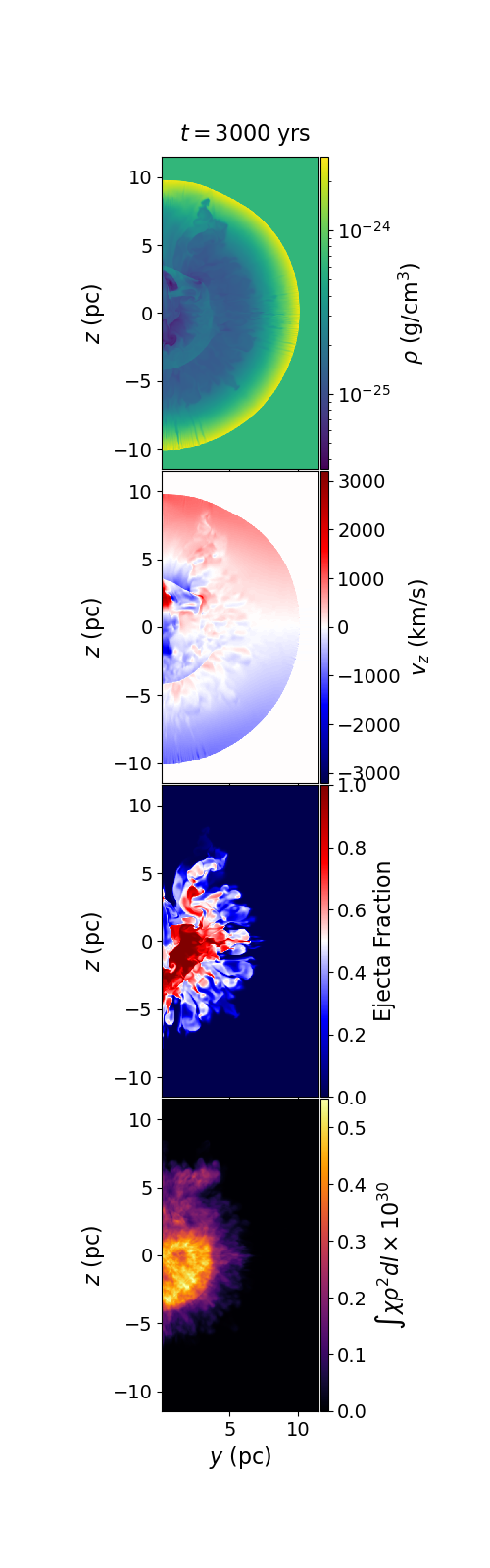} \\
\end{tabular}
\caption{Density, $z$-velocity, ejecta fraction, and X-ray emission measure for the fiducial DD model.
\label{fig:gridathena}}
\end{figure*}

\begin{figure*}
\begin{tabular}{llll}
\includegraphics[height=1.1\linewidth,trim={2cm 0 2cm 0},clip]{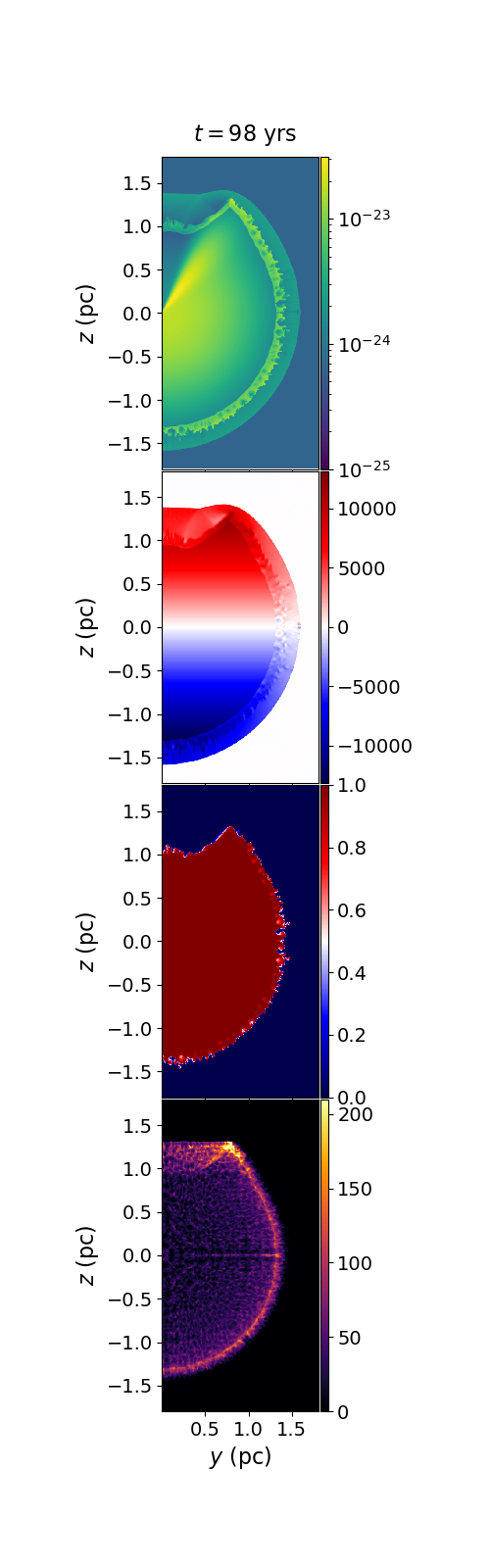} & 
\includegraphics[height=1.1\linewidth,trim={3cm 0 2cm 0},clip]{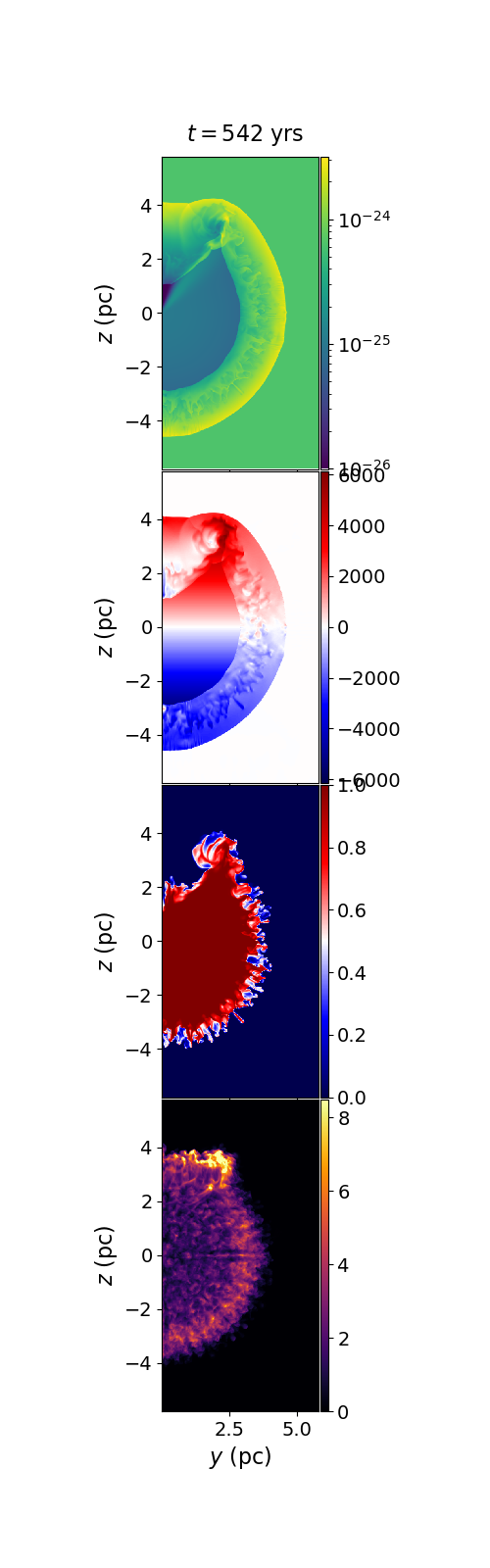} &
\includegraphics[height=1.1\linewidth,trim={3cm 0 2cm 0},clip]{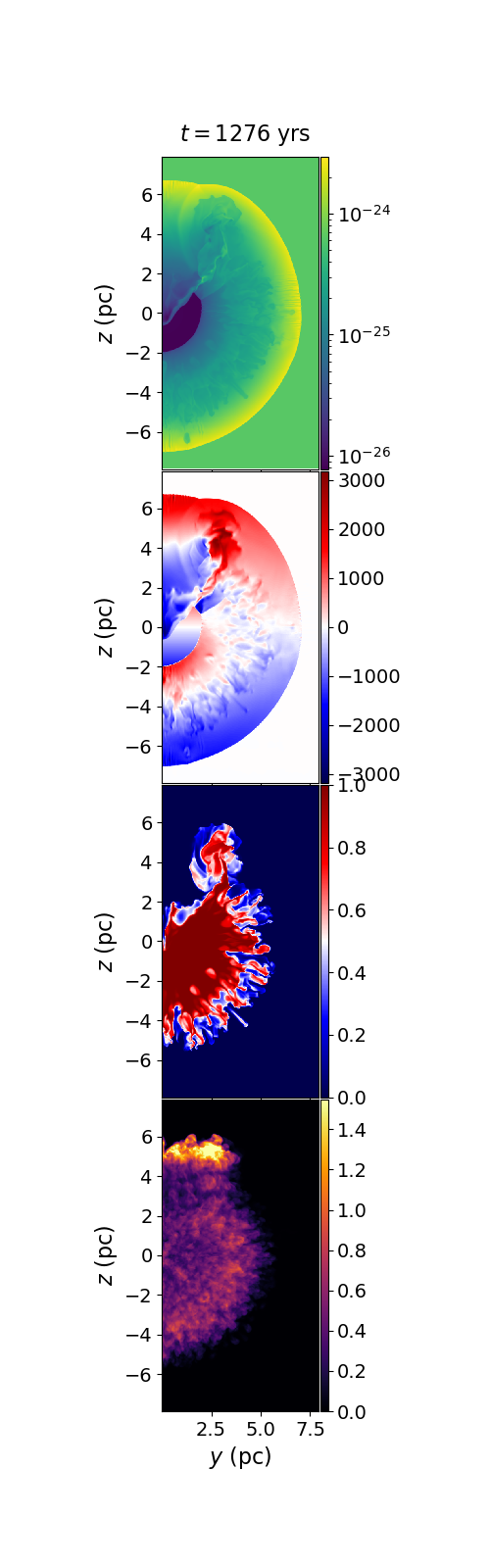} &
\includegraphics[height=1.1\linewidth,trim={3cm 0 1.5cm 0},clip]{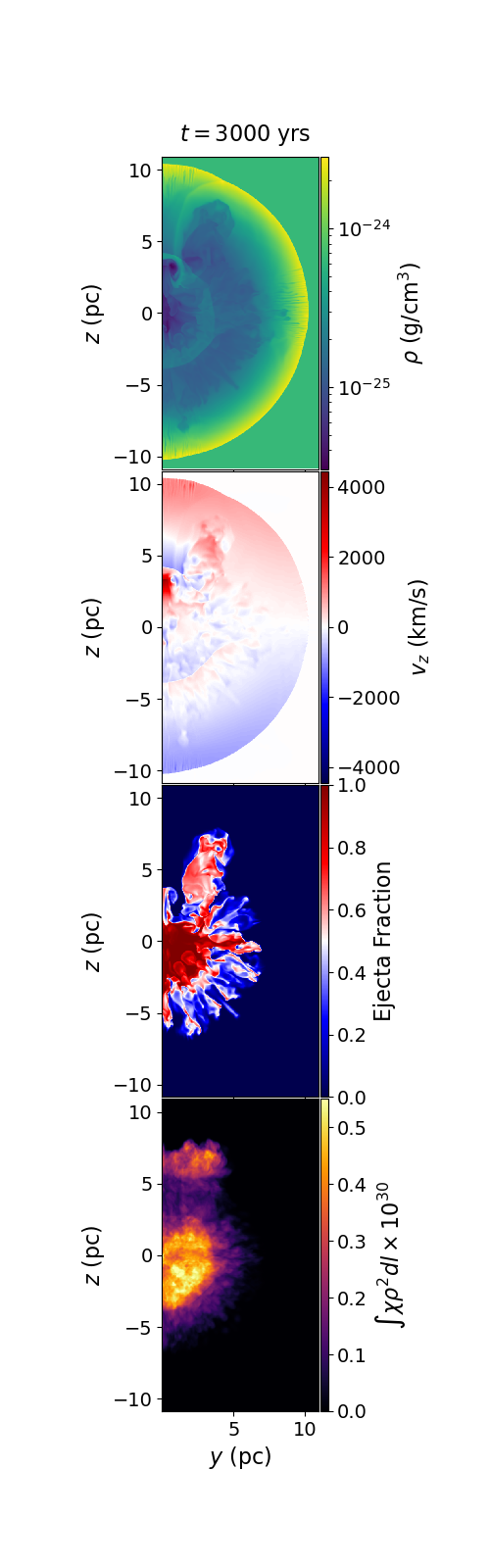} \\
\end{tabular}
\caption{Density, $z$-velocity, ejecta fraction, and X-ray emission measure for the SD model.
\label{fig:gridkasen}}
\end{figure*}


Using the setup described above, we simulate the evolution of the SNR from 10 to 3000 yr for the DD (Fig.~\ref{fig:gridathena}) and SD (Fig.~\ref{fig:gridkasen}) cases. These figures show the gas density, $z$-velocity, ejecta mass fraction $\chi$, and emission measure ($\int\chi\rho^{2}dl$ along a line of sight). The emission measure is a proxy for the thermal X-ray emission and is discussed further in section \ref{sec:tomography}. We find that the trajectories of both the forward and reverse shocks and the composition of the SNR are greatly altered within the wake, each of which is discussed in detail below.

\subsection{Forward Shock} \label{sec:forwardshock}

\begin{figure}
    \centering
    \includegraphics[width=1.0\linewidth]{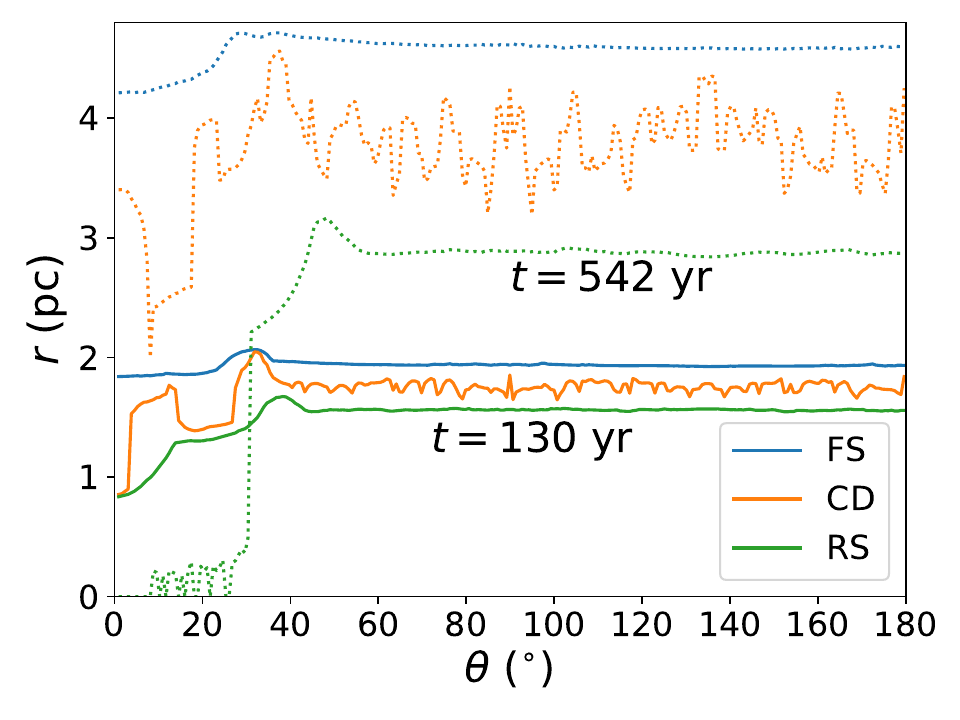}
    \caption{Shock surfaces and contact discontinuity (defined here as $\chi=0.5$) at $t=130$ yr (solid lines) and $t=542$ yr (dotted lines) in the DD model. At both times, the indentation of the FS and protrusion of the CD toward the FS are visible.}
    \label{fig:surfaces}
\end{figure}

In both the DD and SD models, the forward shock within the wake quickly falls behind that of the unperturbed ejecta. Although the DD model initially contained high-velocity material accelerated by the companion interaction, this gas was quickly slowed by the ISM. Similarly, the forward shock slightly protrudes ahead of the unperturbed ejecta at the edge of the wake ($\theta\approx 30$ -- $35^{\circ}$) where the density is highest. In the DD model the protrusion of the FS is narrower in its angular extent, reflecting the differences in ejecta structure between the two cases (Fig.~\ref{fig:kasenvsathena}). We show the shape of the FS at $t=130$ yr and $t=542$ yr in Fig.~\ref{fig:surfaces} (as well as the reverse shock and contact discontinuity).

Interestingly, at $t\approx1000$ yrs the forward shock within the wake catches up to that of the unperturbed ejecta, and after this point the FS is roughly spherical. This is because pressure gradients in the shocked ejecta are large enough to induce transverse flow, forcing gas into the wake.


\subsection{Reverse Shock} \label{sec:reverseshock}

Due to the low density within the wake, the reverse shock travels there much faster than the unperturbed ejecta and more effectively reverses the direction of the fluid velocity. It then passes through the center of the remnant and collides with the reverse shock traveling the other direction at a location $\approx$2 pc from the origin. Following this, the RS ``bounces'' outward, adding additional heat to the center of the SNR. Because of the asymmetrical shape of the bounce shock, it is particularly strong along the $+z$-direction. This forces gas into the wake at high velocity (peaking at $\sim$10,000 km/s) before being slowed by the large negative radial velocity of the reverse-shocked material present within the wake. This effect is hampered in the DD model by the particularly low density along the axis of symmetry, which allows the reverse shock to more quickly pass the center of the remnant and converge farther off-center.

\subsection{Contact Discontinuity} \label{sec:cd}

The shape of the contact discontinuity largely follows that of the forward shock: protruding at the edge of the wake and indented within the wake. However, while the forward shock becomes spherical at late times, the contact discontinuity continues to become increasingly asymmetrical. Here we define the contact discontinuity as the $\chi=0.5$ surface, where $\chi$ is the mass fraction of ejecta, though full slice plots of $\chi$ are shown in the third row of Figs.~\ref{fig:gridathena} and \ref{fig:gridkasen}. We see that the differential flow at the edge of the wake -- caused by the reverse shock passing through the wake -- creates a large vortex ring encircling the wake. This suspends a torus of ejecta at larger radii than that of the normal Rayleigh-Taylor plumes. The vortex is more pronounced in the SD model but is present in both, and plays a large role in determining the X-ray morpohology of the SNR (see section \ref{sec:tomography}). Another consequence of the RS quickly traversing the wake is that it draws ISM deep into the remnant through the wake. In both ejecta models, ISM material eventually reaches the center of the remnant!

\subsection{Comparison to Theory} \label{sec:theory}


\begin{figure}
    \centering
    \includegraphics[width=1.0\linewidth]{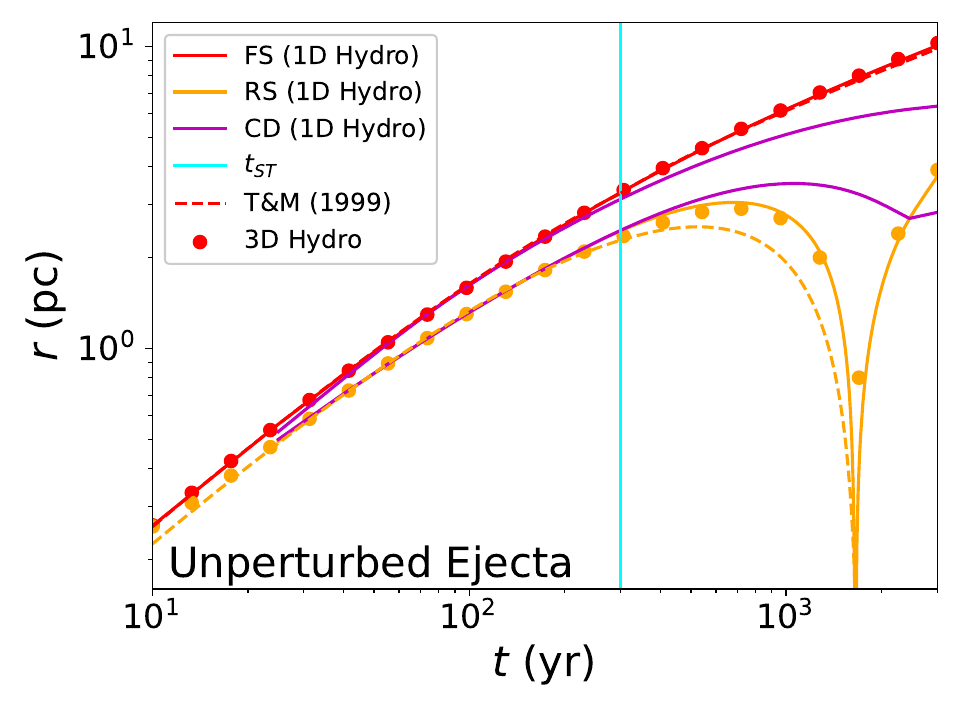}
    \caption{Comparison between our 1-D (solid lines), 3-D (dots), and semi-analytic (dashed lines) predictions for the shock trajectories within the unperturbed ejecta, as well as the transition from the ED phase to the ST phase (blue line). The two magenta lines designate the maximum and minimum extent of RT stirring.}
    \label{fig:theory90}
\end{figure}

To analyze our 3-D results, it is useful to compare them to 1-D models of SNR evolution. To this end, we compute semi-analytic predictions of shock trajectories assuming spherical symmetry. We also perform 1-D hydrodynamical simulations of the SNR along specific radial rays, assuming no transverse flow. 

Analytical models of shock trajectories in spherically-symmetric supernova remnants were pioneered by \citet{1999ApJS..120..299T} assuming broken power-law density profiles. Here the density was expressed in terms of a ``structure function'' $f(w)$:
\be
\rho(r,t)=\frac{\Mej}{\vmax^{3}t^{3}}f(w),
\ee
where $w \equiv v/\vmax$. We extended the formalism of \citet{1999ApJS..120..299T} to the Gaussian profile (\ref{eq:densityProfileVelocity}) by choosing the structure function to be
\be
f(w) &=& \left(\frac{\vmax}{v_{0}\sqrt{\pi}}\right)^{3}\exp\left(-\frac{\vmax^{2}}{v_{0}^{2}}w^{2}\right) \\
&=& \left(\frac{3}{2\pi\alpha}\right)^{3/2}\exp\left(-\frac{3}{2\alpha}w^{2}\right). \label{eq:structurefunction}
\ee
Here $\alpha\equiv \Eej/(\Mej\vmax^{2}/2)$. \citet{1999ApJS..120..299T} split the remnant evolution into an ejecta-dominated (ED) phase and a Sedov-Taylor (ST) phase, with different models for the shock evolution in each. In particular, they assumed that the ``lead factor'' $l$ (ratio of FS radius to RS radius) and the ratio of the post-shock pressures of the FS and RS $\varphi$ are both constant during the ED phase. That is, $\varphi(t)=\varphi(0)=\varphi_{\rm ED}$ and $l(t)=l(0)=l_{\rm ED}$. These are free parameters which we choose to be $l_{\rm ED}=1.1$ and $\varphi_{\rm ED}=0.5$. Given these assumptions, the ED evolution of the FS radius is then given by
\be
R_{\rm FS}^{3/2}=\frac{3}{2}\sqrt{\frac{l_{\rm ED}\Mej}{\varphi_{\rm ED}\rho_{\rm ISM}}}\int_{R_{\rm FS}/\vmax t l_{\rm ED}}^{1}dw\sqrt{wf(w)}.
\ee
Using the Gaussian structure function (\ref{eq:structurefunction}), this evaluates to 
\be
\begin{aligned}
R_{\rm FS}^{3/2} &= \frac{3}{2^{5/4}\pi^{3/4}} \sqrt{\frac{l_{\rm ED}\Mej}{\varphi_{\rm ED}\rho_{\rm ISM}}} \\ &\times\left[\Gamma\left(\frac{3}{4},\frac{3}{4\alpha}\left[\frac{R_{\rm FS}}{l_{\rm ED}\vmax t}\right]^{2}\right)-\Gamma\left(\frac{3}{4},\frac{3}{4\alpha}\right)\right],
\end{aligned}
\ee
where $\Gamma$ is the upper incomplete gamma function. This implicit equation can be solved numerically for $R_{\rm FS}$. With this in hand, the RS evolution follows trivially given a constant lead factor.

In \citet{1999ApJS..120..299T}, during the ST phase the FS obeys the familiar solution for a strong blast wave \citep{1950RSPSA.201..159T} and the reverse shock is assumed to have a constant acceleration relative to the oncoming ejecta $\tilde{a}_{\rm RS}$. This is a free parameter for which we choose 
\be
\tilde{a}_{\rm RS} = -0.12 \Eej \Mej^{-4/3} \rho_{\rm ISM}^{1/3}.
\ee
This formalism requires a choice of the epoch of transition between the ED and ST phases $t_{\rm ST}$. We choose $t_{\rm ST}$ to be the time at which the ISM mass swept up by the forward shock is equal to $\Mej$ assuming constant expansion velocity:
\be
t_{\rm ST} = \frac{1}{\vmax}\left(\frac{3}{4\pi}\frac{\Mej}{\rho_{\rm ISM}}\right)^{1/3}.
\ee
These choices yield FS and RS trajectories which can be compared to those of our simulations.

We also perform 1-D hydrodynamical simulations of the SNR using RT1D \citep{rt1d}, a 1-D code which has been calibrated to include the effects of RT instabilities based on the results of 3-D simulations. These simulations ignore any transverse flow and serve to validate the semi-analytic predictions above. We perform two simulations: one for the unperturbed ejecta and the other along the center of the wake. As discussed in section \ref{sec:kasen}, these ejecta profiles differ only by a constant factor (0.1) in the SD case. The SD model is chosen for these calculations as the ejecta structure is independent of $\theta$ in the interior of the wake (Fig.~\ref{fig:kasenvsathena}), whereas in the DD case there is a large $\theta$ dependence. 

We compare our 3-D, 1-D, and semi-analytic models for the unperturbed ejecta in Fig.~\ref{fig:theory90}. Here the two magenta lines enclose the region in which RT stirring occurs. We see that all three models agree on the FS position for the duration of the run. Both hydrodynamical simulations also agree on the reverse shock evolution -- including the convergence time -- though the semi-analytic model deviates from the numerics during the ST phase. This is due to the assumption of constant acceleration, which breaks down at the center of the remnant where the acceleration is much larger. The semi-analytic model still gives the correct convergence time, but only due to our choice of $\tilde{a}_{\rm RS}$.

\subsection{Expectations within the Wake} \label{sec:wakeexpectations}

\begin{figure}
    \centering
    \includegraphics[width=1.0\linewidth]{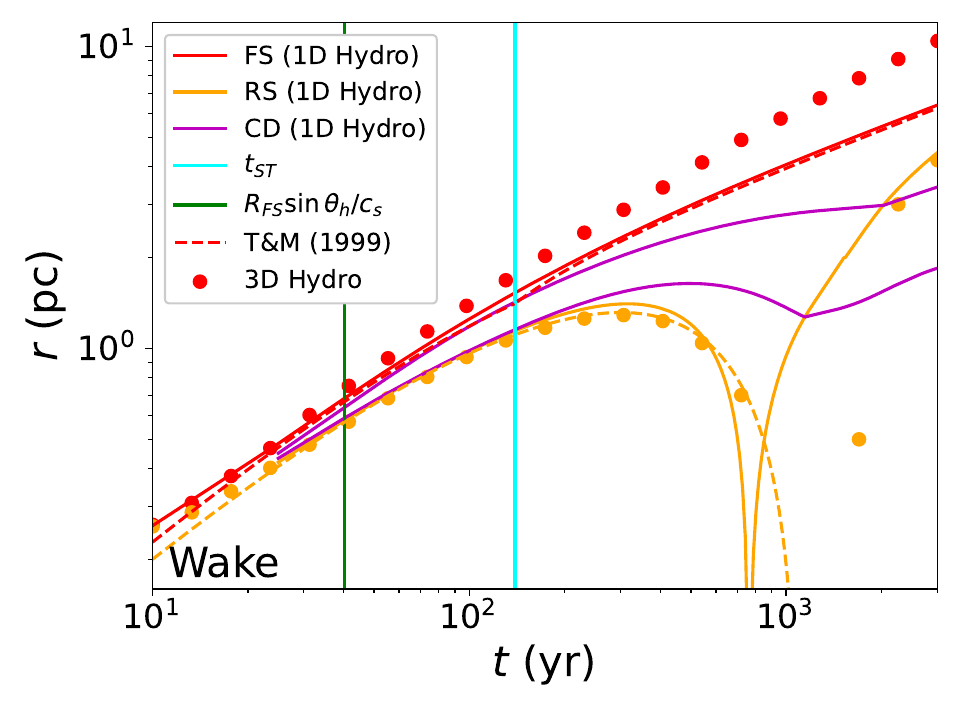}
    \caption{Comparison between our 1-D (solid lines), 3-D (dots), and semi-analytic (dashed lines) predictions for the shock trajectories within the wake of the SD model. Also shown are the transition from the ED phase to the ST phase (blue line) and the sound-crossing time of the wake (green line).}
    \label{fig:theory0}
\end{figure}

We now turn to the shock trajectories within the wake (Fig.~\ref{fig:theory0}). For the semi-analytic model, the evolution within the wake is obtained by scaling $\Mej$ and $\Eej$ by 0.1. This also results in an earlier transition to the ST phase (blue line). Here the 1-D and semi-analytic models agree on the FS location, but the 3-D model deviates from these after only a few tens of years. This is comparable to the sound-crossing time of the wake (green line), indicating that the deviation is due to transverse flow driven by pressure gradients in the shocked ejecta and providing a mechanism by which the FS is able to regain spherical symmetry. The RS evolution is similar in the 1-D and 3-D models before the RS convergence but differs afterward. This is unsurprising as the 1-D model is unable to include the off-center convergence described in \ref{sec:reverseshock}.

\section{X-Ray Tomography} \label{sec:tomography}


\begin{figure*}[p!]
\begin{tabular}{lll}
\includegraphics[height=0.48\textwidth,trim={1cm 0 0.5cm 0},clip]{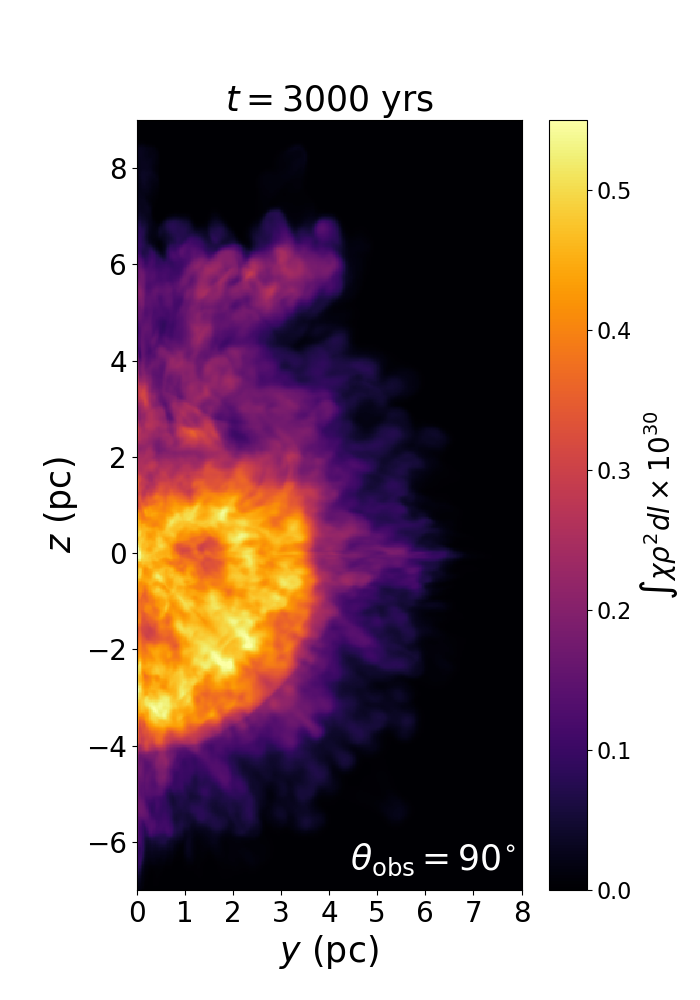} & 
\includegraphics[height=0.48\textwidth,trim={2cm 0 0 0},clip]{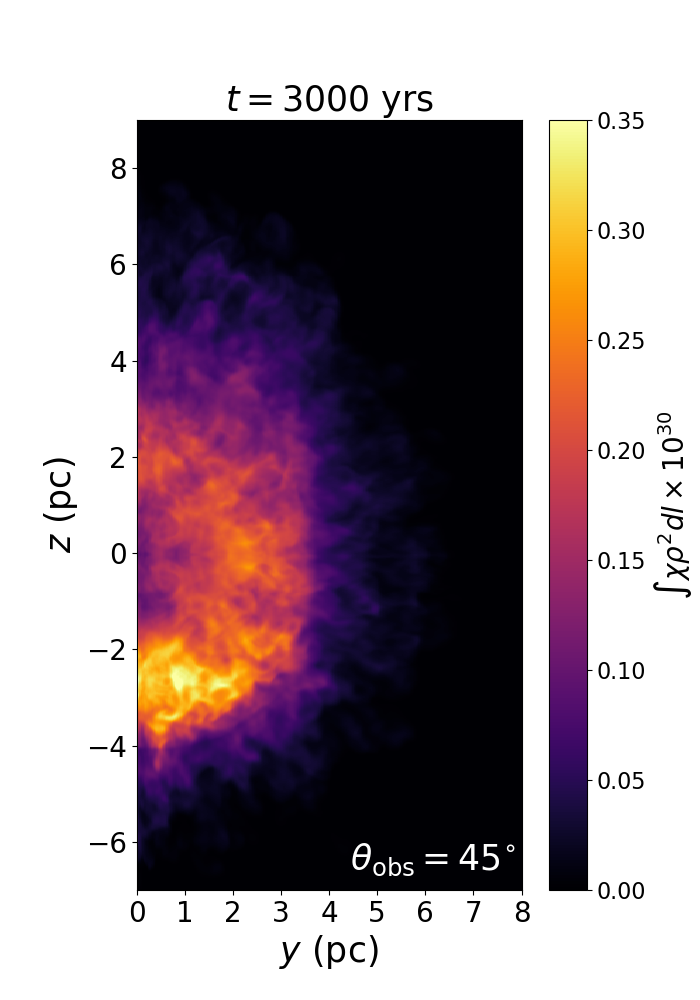} &
\includegraphics[height=0.48\textwidth,trim={2cm 0 0.5cm 0},clip]{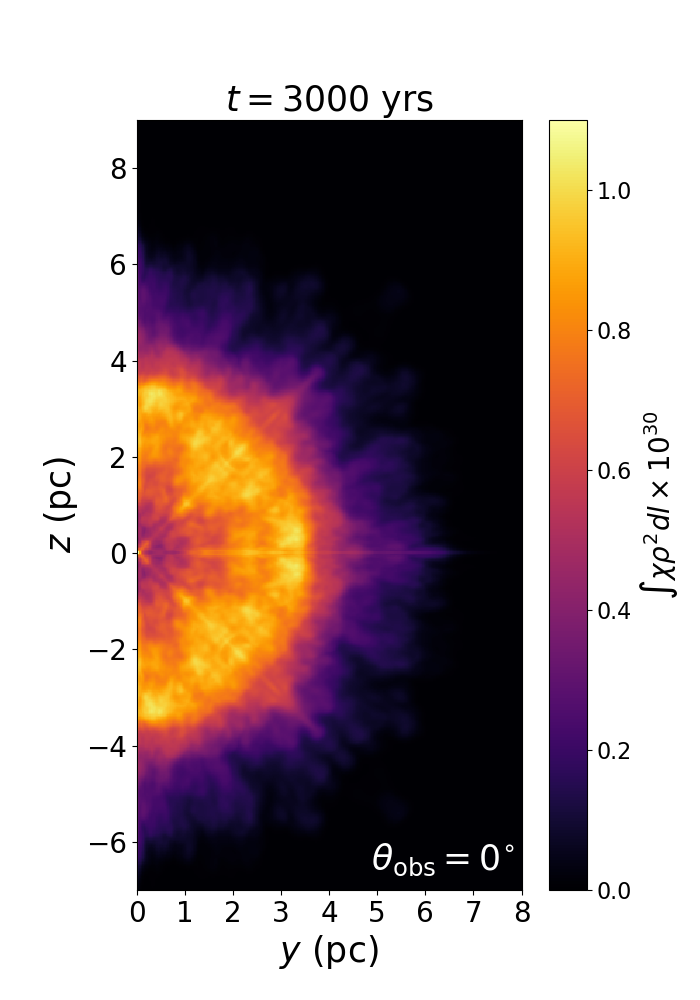} \\
\end{tabular}
\caption{Emission measure for the DD model at $\theta_{\rm obs}=90^{\circ}$ (left), $\theta_{\rm obs}=45^{\circ}$ (center), and $\theta_{\rm obs}=0^{\circ}$ (right). For reference, at this time the FS has a radius of 10 pc.
\label{fig:bremangles}}
\end{figure*}

\begin{figure*}
\begin{tabular}{lll}
  \includegraphics[height=0.48\textwidth,trim={1cm 0 0.5cm 0},clip]{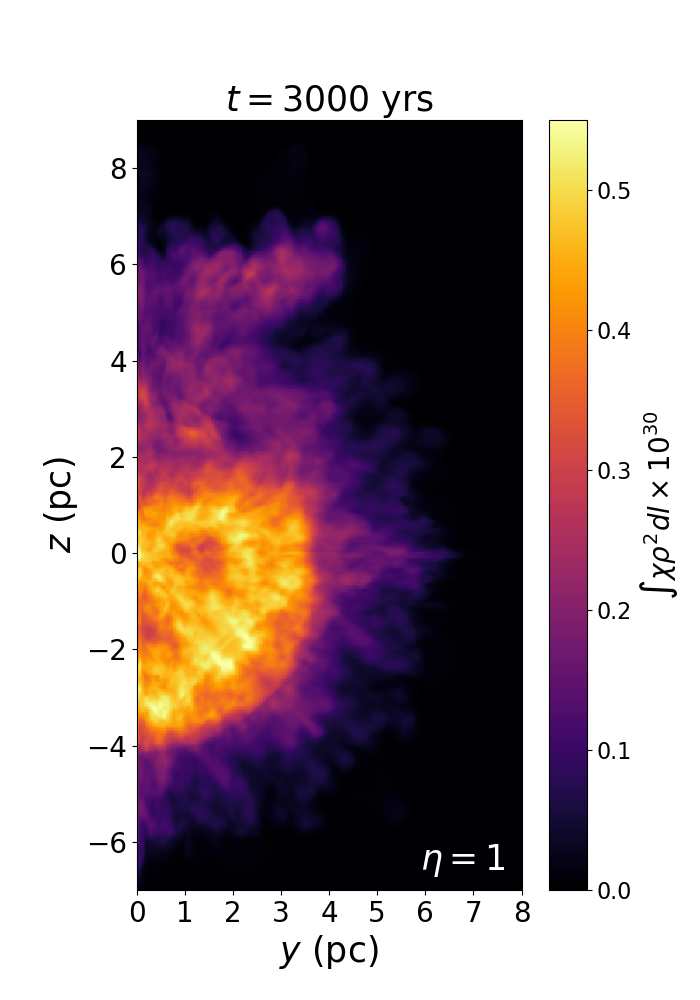} & 
  \includegraphics[height=0.48\textwidth,trim={2cm 0 0 0},clip]{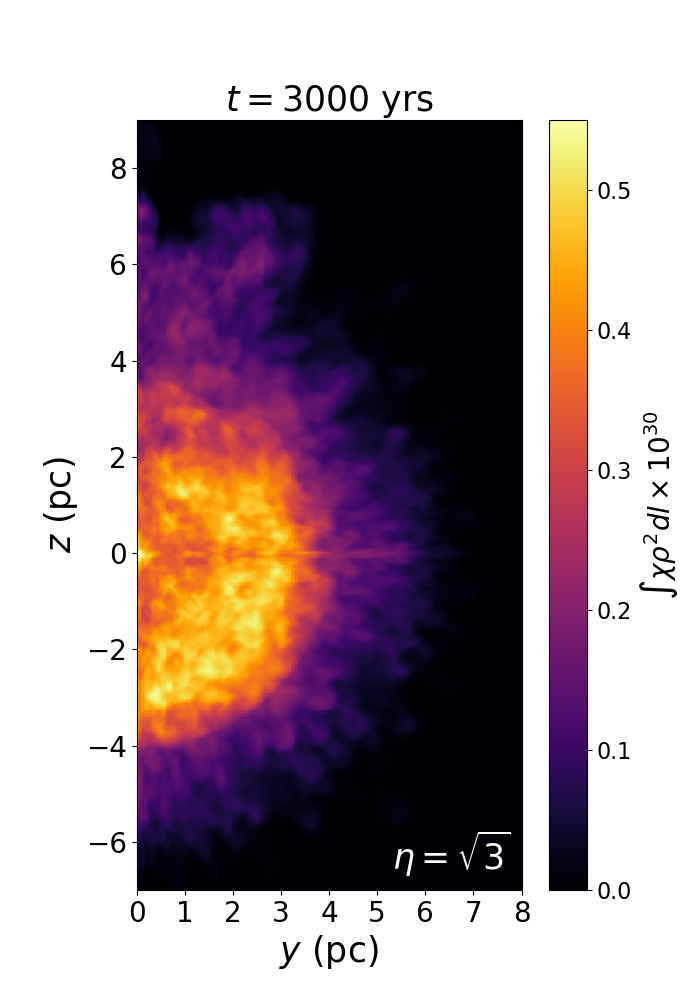} &
  \includegraphics[height=0.48\textwidth,trim={2cm 0 0.5cm 0},clip]{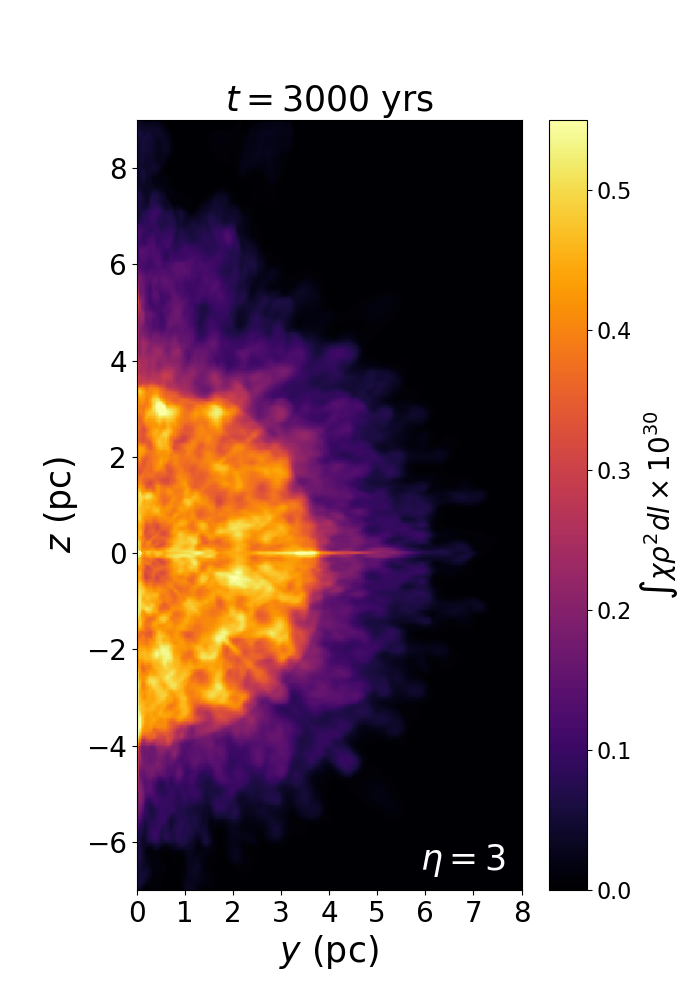} \\
\end{tabular}
\caption{X-ray emission measures for our Athena++ models with orbital separation increasing to the right: $\eta=1$ (left), $\eta=\sqrt{3}$ (center), and $\eta=3$ (right). As the orbital separation increases, the X-ray morphology becomes spherical, as expected. The viewing angle is $\theta_{\rm obs}=90^{\circ}$ for all panels.
\label{fig:bremathena}}
\end{figure*}


Because SNRs are optically thin to X-rays, we estimate the thermal bremsstrahlung emission by integrating $\rho^{2}$ along a given line of sight. We emphasize that this captures only the continuum free-free emission, neglecting line emission. For a comprehensive review of SNR X-ray radiation mechanisms, see \citet{2012A&ARv..20...49V}.

However, X-ray spectroscopy is excellent at discerning the composition of a remnant, which our methodology is suited to predict. As we showed in section \ref{sec:cd}, the distribution of ejecta within the SNR is asymmetrical and contains a toroidal structure around the wake. This manifests in the emission measure $\int\chi\rho^{2}dl$ -- where $\chi$ is the mass fraction of ejecta -- which is shown in the bottom rows of Figs.~\ref{fig:gridathena} and \ref{fig:gridkasen}.
The flat edge seen by \citet{2016ApJ...833...62G} at $t=100$ and 300 yrs (see their Fig.~10) is also present here. 

We also show the DD model from three different viewing angles $\theta_{\rm obs}$ in Fig.~\ref{fig:bremangles}. The asymmetry is most pronounced when the line of sight is normal to the axis of symmetry (left panel), but is still visible from other angles (center panel) and differs somewhat from that of a spherical remnant even when looking directly into the wake (right panel).

As discussed in section \ref{sec:athenasetup}, we have also performed simulations in which the orbital separation is increased by a factor $\eta$ without changing the donor size to explore the limiting case of a small solid angle subtended by the donor. The emission measure for each $\eta$ value is shown in Fig.~\ref{fig:bremathena} for $\theta_{\rm obs}=90^{\circ}$. As the orbital separation increases, the morphology becomes increasingly spherical, as one would expect. This demonstrates that our methodology does not inherently lead to predictions of asymmetrical X-ray emission and produces spherical morphology in the appropriate limit.

\section{Discussion \& Conclusions} \label{sec:discussion}

We have computed the evolution of type Ia supernova remnants following a companion interaction event in both the DD and SD cases. The ejecta structure in the DD case was obtained via 3-D hydrodynamical simulations of the ejecta-companion collision in which the companion was treated as a rigid sphere, whereas in the SD case it was obtained from the fit performed by \citet{2010ApJ...708.1025K} to their results following a collision with a red giant donor. We identified several differences between the ejecta structures such as the presence of high-velocity, radiation pressure-dominated ejecta within the wake and a sharply-defined bow shock in the DD case. This provides a possible explanation for the two distinct ejecta velocities (12,400 km/s and 23,500 km/s) found in observations of SN 2021aefx \citep{2023ApJ...959..132N}, though many SNIa show no evidence of companion interaction at early times \citep{2015Natur.521..332O}. Acceleration of ejecta due to companion interaction has been seen to some degree in previous work \citep[e.g.][]{2012ApJ...745...75G,2015MNRAS.449..942P,2017MNRAS.465.2060B,2024arXiv240800125W}.

Both ejecta models are then scaled to $t=10$ yrs and evolved to $t=3000$ yrs using the expanding-grid code Sprout \citep{sprout}. These simulations revealed that both the forward and reverse shocks are altered significantly by the presence of the wake. The forward shock is initially indented, but becomes spherical after $\approx$1000 yrs due to transverse flow of the ejecta to fill in the wake. This is substantiated by 1-D calculations which -- in addition to validating the 3-D results -- demonstrate that the wake fills in on a timescale comparable to its sound-crossing time. The initially high-velocity material within the wake does little to alter the features stated above, as it is quickly slowed by the ISM. The FS morphology found in this paper bears a strong resemblance to the axisymmetric SPH simulations performed by \citet{2012ApJ...745...75G} -- see their Figs.~10 and 15. They treated the wake as both an empty hole and as a region partially filled by material stripped from the (non-degenerate) donor. \citet{2022ApJ...930...92F} included ejecta from the He shell detonation preceeding the SN, which had the highest velocity in a direction opposite to the point of ignition. This caused a protrusion misaligned with the wake that was present for the first few hundred years. Additionally, \citet{2022ApJ...930...92F} included the relative velocity of the SNR relative to the ISM due to the orbital velocity prior to the SN. This resulted in a dipole component to the FS and RS but likely had little effect on the wake as the relative velocity was roughly orthogonal to the wake.

The reverse shock encounters less impedance in the wake and reaches the center of the remnant far more quickly than from any other direction. This results in an off-center convergence of the reverse shock which drives a plume of material back into the wake, leaving the remnant with an asymmetrical core. Though the details of these features vary between the SD and DD models, they are present in both. The models of \citet{2022ApJ...930...92F} and model E of \citet{2012ApJ...745...75G} contained a dense clump of ejecta near the base of the wake, hindering the RS near the center of the remnant and causing the RS to be more symmetrical at late times.

The RS also draws ISM material deep into the wake to eventually reach the center, whereas around the edge of the wake a ring vortex suspends a significant amount of ejecta at large radii. \citet{2012ApJ...745...75G} similarly found large Rayleigh-Taylor (RT) plumes at the edge of the wake, which was corroborated by \citet{2022ApJ...930...92F} using a different numerical scheme. By estimating the continuum thermal emission using a projection of $\rho^{2}$, we obtain predictions of the SNR X-ray morphology. For the first few centuries, the remnant resembles a sphere with a flat edge, in agreement with \citet{2016ApJ...833...62G}. At later times, the asymmetry becomes more pronounced as the unperturbed ejecta falls behind the ring vortex. 


Although many SNIa remnants do exhibit asymmetries, the sample size is limited and it is unclear for which SNRs companion interaction may be responsible. For example, asymmetries in the Tycho remnant are well-explained by a gradient in the ISM \citep{2016ApJ...823L..32W}. For at least a fraction of these remnants, it is likely that both objects were destroyed due to the lack of an observable surviving companion \citep{2023ApJ...950L..10S}. In such cases, the ejecta mass is not limited by the Chandrasekhar mass, though the ejecta mass is difficult to directly measure due to uncertainty in the ISM density. For example, the SNIa remnant 0509-67.5 is consistent with both Chandrasekhar-mass \citep{2022ApJ...938..121A} and sub-Chandrasekhar mass \citep{2019PhRvL.123d1101S} models. Though the destruction of both objects precludes the formation of a low-density wake, the ejecta structure has been shown to be initially asymmetrical \citep{2024arXiv240108011B}. It is unclear to what degree this carries over to the remnant phase, though we leave this question to future work.









\begin{acknowledgments}
We thank Chris White for assistance with the Athena++ EOS, Sunny Wong for useful discussions about the response of the donor to the supernova, and Soham Mandal for helpful pointers regarding the use of Sprout. This research benefited from interactions with a variety of researchers that were funded by the Gordon and Betty Moore Foundation through Grant GBMF5076. Computational resources for this work were provided by the Expanse supercomputer at the San Diego Supercomputer Center through allocation PHY240019 from the Advanced Cyberinfrastructure Coordination Ecosystem: Services \& Support (ACCESS) program \citep{access}. ACCESS is supported by National Science Foundation grants \#2138259, \#2138286, \#2138307, \#2137603, and \#2138296. This research was supported in part by grant NSF PHY-2309135 to the Kavli Institute for Theoretical Physics (KITP). LJP is supported by a grant from the Simons Foundation (216179, LB). We use the Matplotlib \citep{Hunter:2007} and SciPy \citep{2020SciPy-NMeth} software packages for the generation of plots in this paper. Use was made of computational facilities purchased with funds from the National Science Foundation (CNS-1725797) and administered by the Center for Scientific Computing (CSC). The CSC is supported by the California NanoSystems Institute and the Materials Research Science and Engineering Center (MRSEC; NSF DMR 2308708) at UC Santa Barbara.
\end{acknowledgments}

\software{Athena++ \citep{athena++},
          Matplotlib \citep{Hunter:2007},
          NumPy \citep{harris2020arrayNUMPY},
          RT1D \citep{rt1d},
          SciPy \citep{2020SciPy-NMeth},
          Sprout \citep{sprout}
          }

\newpage

\appendix

\section{Fitting Functions for Ejecta Structure} \label{sec:fits}

\begin{figure}
    \centering
    \includegraphics[width=1.0\linewidth]{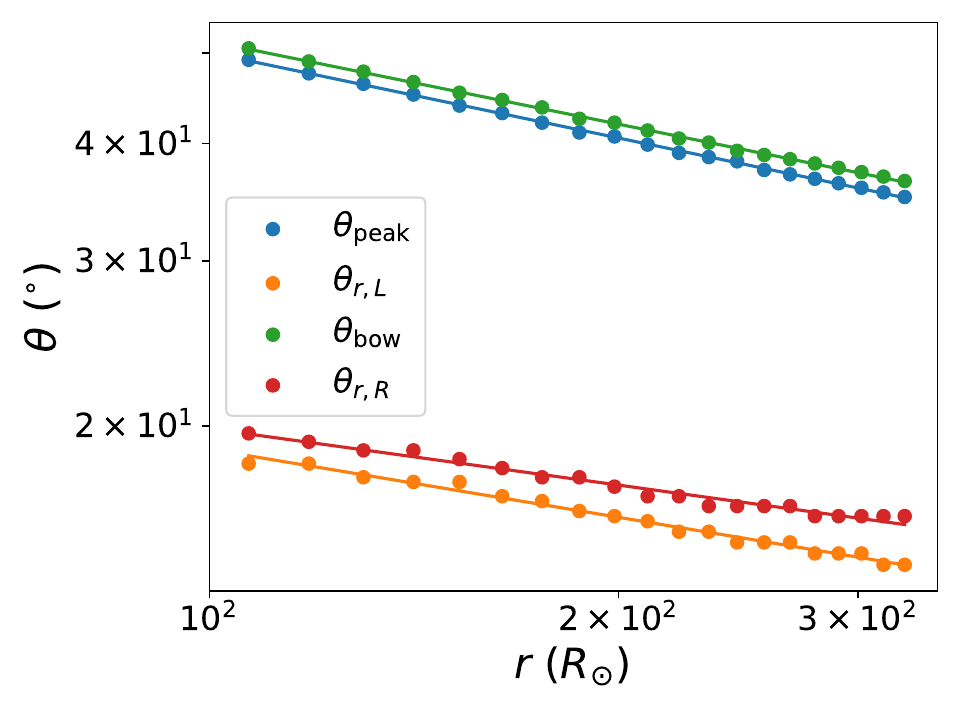}
    \caption{Positions of important features in the density profiles at $t=10000$ s in our fiducial Athena++ model, all of which can be fit to power laws (solid lines).}
    \label{fig:fits}
\end{figure}

To map our Athena++ data into Sprout, we fit the density structure, the outer edge of the ejecta, and the shape of the bow and recompression shocks to analytical functions. This allows us to extend our Athena++ data -- which had limited extent in both the radial and polar directions -- to all $r$ and $\theta$. It also ensures smoothness of the fluid, which assists with numerical stability.

We use a different fitting function for each region: ejecta which passed through both shocks ($f_{1}$), ejecta which passed through only the bow shock ($f_{3}$), the transition between these two regions at the recompression shock ($f_{2}$), and unperturbed ejecta. Here the density at a given radius is normalized by that of the unperturbed ejecta $\rho_{G}(r)$ at that radius:
\be
\rho(r,\theta)/\rho_{G}(r) = \begin{cases} f_{1}(\theta) & 0 < \theta < \theta_{r,L}, \\
f_{2}(\theta) & \theta_{r,L} < \theta < \theta_{r,R}, \\
f_{3}(\theta) & \theta_{r,R} < \theta < \theta_{\rm bow}, \\
\max[1,f_{3}(\theta)] & \theta_{\rm peak} < \theta < \theta_{\rm bow}, \\
1 & \min(\theta_{\rm bow},\pi/2)<\theta. \end{cases}
\ee
These fitting functions are given by:
\begin{align}
\begin{split}
f_{1} ={}& \exp\left(C_{1}+C_{2}\ln\left(\frac{\theta}{\theta_{r,L}}\right)\right.\\&+\left.C_{3}\left[\ln\left(\frac{\theta}{\theta_{r,L}}\right)\right]^{2}\right), 
\end{split} \\
\begin{split}
f_{2} ={}& \left(\tanh\left[-6\frac{\theta-\theta_{r,L}}{\theta_{r,R}-\theta_{r,L}}+3\right]+1\right)\\&\times\frac{\rho_{\rm max}-\rho_{\rm min}}{2}+\rho_{\rm min}, 
\end{split} \\
\begin{split}
f_{3} ={}& C_{4}+20\sin\left(1.015\pi\frac{\theta}{\theta_{\rm bow}}\right)\left(\frac{\theta}{\theta_{\rm peak}}\right)^{16C_{5}}\\&+C_{5}\left(\frac{\theta}{\theta_{\rm peak}}\right)^{2.75C_{5}},
\end{split}
\end{align}
where
\be
\rho_{\rm max}&=&\max[\rho(r,\theta_{r,L}),\rho(r,\theta_{r,R})]/\rho_{G}(r), \\
\rho_{\rm min}&=&\min[\rho(r,\theta_{r,L}),\rho(r,\theta_{r,R})]/\rho_{G}(r).
\ee
We also enforce $\rho(r,\theta)\geq 0.01\rho_{G}(r)$ so that the density does not drop too low at $\theta=0$. The coefficients in these fits are also fit as functions of $r$:
\begin{align}
\begin{split}
C_{1} ={}& -0.46218-3.2527r'\\&+2.2262r'^{2}+0.31r'^{3}-0.060088r'^{4}
\end{split} \\
C_{2} ={}& -0.50302\ln(r)+15.959, \\
C_{3} ={}& -0.062855\ln(r)+1.9343, \\
\begin{split}
C_{4} ={}& 1.0313-3.5916r'\\&+5.7664r'^{2}-4.4343r'^{3}+1.4134r'^{4},
\end{split} \\
C_{5} ={}& \exp(-0.25425\ln r+8.3664).
\end{align}
Here $r' = r/(1~{\rm AU})$. The density profiles resulting from these fits are compared with our Athena++ data in Fig.~\ref{fig:kasenvsathena}. We find that the shape of the shock surfaces can be approximated by power laws: 
\be
\ln\theta_{\rm bow}  &=& -0.29337\ln r+8.5690, \\
\ln\theta_{\rm peak} &=& -0.30071\ln r+8.7564, \\
\ln\theta_{r,L} &=& -0.24092\ln r+6.0168, \\
\ln\theta_{r,R} &=& -0.19975\ln r+4.8498,
\ee
as shown in Fig.~\ref{fig:fits}.
The ejecta surface can be approximated using piecewise polynomial fits in $\theta$. For $\eta=1$,
\be
R(\theta)/R_{0} = \begin{cases} r_{0} + 2.4563\theta^{2} & \theta<\theta_{1}, \\ 
-4.30108\theta+2.430108 & \theta_{1}<\theta<\theta_{2}, \\
-2.20514(\theta-\theta_{2})^{2}+r_{2} & \theta_{2}<\theta<\theta_{3}, \\
1 & \theta_{3}<\theta. \end{cases}
\ee
where $(\theta_{0},\theta_{1},\theta_{2})=(0.205,0.28,0.6$), $(r_{0},r_{1},r_{2})=(1.44516,1.5484,1.2258)$, and $R_{0}$ is the outer surface of unperturbed ejecta at a given time. 

\section{Upwind-Preserving Hydrodynamics on an Expanding Grid} \label{sec:upwind}

The solution to the Riemann problem on a static grid can be adapted to a moving grid using two simple modifications:
\begin{enumerate}
    \item For a cell interface moving with velocity $\facev$, the face flux $\flux_{s}$ is evaluated within the region of the Riemann fan corresponding to the characteristic $\vel=\facev$ rather than $\vel=0$.
    \item Given a fluid state $\state$, an advection term $-\state\facev^{T}$ is added to the face flux to account for the material ``swept up'' by the face as it moves.
\end{enumerate}
The flux across a moving interface is then
\be
\flux_{m} = \flux_{s} - \state \facev^{T}, \label{eq:fluxm}
\ee
or
\be
\flux_{m} = \left( \begin{array}{c} \rho\vel^{T} \\ \rho\vel\vel^{T} + P I \\ (\rho \epsilon + \rho v^{2}/2 + P) \vel^{T} \\ \rho\chi_{i}\vel^{T} \end{array} \right) - \left( \begin{array}{c} \rho \\ \rho\vel \\ \rho e \\ \rho\chi_{i} \end{array} \right) \facev^{T}.
\ee
Here $I$ is an identity matrix and $\chi_{i}$ is the mass fraction of the $i$-th passive tracer. Moving-mesh methods are typically designed such that the fluid velocity is similar to the face velocity, with the goal of maintaining Galilean invariance. A side effect of this is that the two terms on the right side of (\ref{eq:fluxm}) nearly cancel out, i.e.~$|\flux_{m}| \ll |\flux_{s}| \approx |\state\facev^{T}|$. Thus, small numerical errors may cause sign changes in $\flux_{m}$, reversing the direction of the numerical flux and violating the upwind property of the scheme.

This can be rectified by a simple change of inertial frames: we boost into the rest frame of the cell interface, compute the face flux, and finally boost the face flux back into the ``lab'' frame (i.e.~the global rest frame). The flux then takes on a similar form:
\be
\flux_{m} = \flux'_{s} - \flux_{\rm c}. \label{eq:fluxmprime}
\ee
Here $\flux'_{s}$ is the static flux $\flux_{s}$ with the replacement $\vel\rightarrow\vel-\facev$:
\be
\flux'_{s} = \left( \begin{array}{c} \rho(\vel-\facev)^{T} \\ \rho(\vel-\facev)(\vel-\facev)^{T} + P I \\ (\rho\epsilon + \rho(\vel-\facev)^{2}/2 + P) (\vel-\facev)^{T} \\ \rho\chi_{i}(\vel-\facev)^{T} \end{array} \right).
\ee
It is useful at this point to define the static flux normal to the face as
\be
\flux'_{s}\cdot\normal = \left( \begin{array}{c} Q_{1} \\ \qflux_{2} \\ Q_{3} \\ Q_{4} \end{array} \right),
\ee
as these fluxes will soon come in handy. The flux correction $\flux_{\rm c}$ is the equivalent of the advection term but takes on the form
\be
\flux_{\rm c} = \left( \begin{array}{c} 0 \\ -\rho(\vel-\facev)\facev^{T} \\ \rho w^{2}(\vel-\facev)/2 - \rho(\vel\cdot\facev)(\vel-\facev) - P\facev \\ 0 \end{array} \right). \label{eq:fcorr}
\ee
If we project along the face normal $\normal$, the flux correction can be written succinctly as
\be
\flux_{\rm c}\cdot\normal = \left( \begin{array}{c} 0 \\ -Q_{1}\facev^{T} \\ -\qflux_{2}\cdot\facev-w^{2}Q_{1}/2 \\ 0 \end{array} \right).
\ee

At the end of each timestep, the fluid states are updated using these fluxes in conjunction with the cell volumes and face areas. On unstructured grids such as Voronoi tessellations, these geometric properties of the grid must be measured at discrete times. However, on a globally-expanding cubical grid, \citet{sprout} showed that the changing grid geometry can be directly incorporated into the flux computation. Here coefficients are introduced to account for the change in the volume of the cells ($C_{U}$) and in the area of the cell interfaces ($C_{F}$):
\be
\flux_{m} = C_{F} \flux_{s} - C_{U} \state\facev^{T}.
\ee
Thus, to write the flux in the form of (\ref{eq:fluxmprime}), we must recompute the flux correction. By equating 
\be
\flux_{m} = C_{F} \flux_{s} - C_{U} \state\facev^{T} = C_{F} \flux'_{s} - \flux'_{\rm c},
\ee
we obtain 
\be
\begin{aligned}
\flux'_{\rm c}\cdot\normal = C_{F}\left( \begin{array}{c} 0 \\ -Q_{1}\facev \\ -\qflux_{2}\cdot\facev-w^{2}Q_{1}/2 \\ 0 \end{array} \right) \\ - (C_{U}-C_{F})\left( \begin{array}{c} \rho w \\ \rho\vel w \\ \rho e w \\ \rho\chi_{i} w \end{array} \right),
\end{aligned}
\ee
or
\be
\flux'_{\rm c}\cdot\normal = C_{F} \flux_{\rm c}\cdot\normal - (C_{U}-C_{F}) \state w.
\ee
We find that the flux correction on an expanding grid can be separated into two terms: one which is proportional to the flux correction on a moving but non-expanding grid and one which is proportional to the advection term on a static grid. Note that if the grid is not expanding, $C_{U}=C_{F}=1$ and we recover $\flux'_{\rm c}\cdot\normal=\flux_{\rm c}\cdot\normal$. Putting this all together, the face flux on an expanding grid which preserves the upwind property is
\be
\begin{aligned}
\flux_{m}\cdot\normal = C_{F} \left( \begin{array}{c} Q_{1} \\ \qflux_{2}+Q_{1}\facev \\ Q_{3}+\qflux_{2}\cdot\facev+w^{2}Q_{1}/2 \\ Q_{4} \end{array} \right) \\ + (C_{U}-C_{F})\state w.
\end{aligned}
\ee

\begin{figure}
  \includegraphics[width=1.0\linewidth]{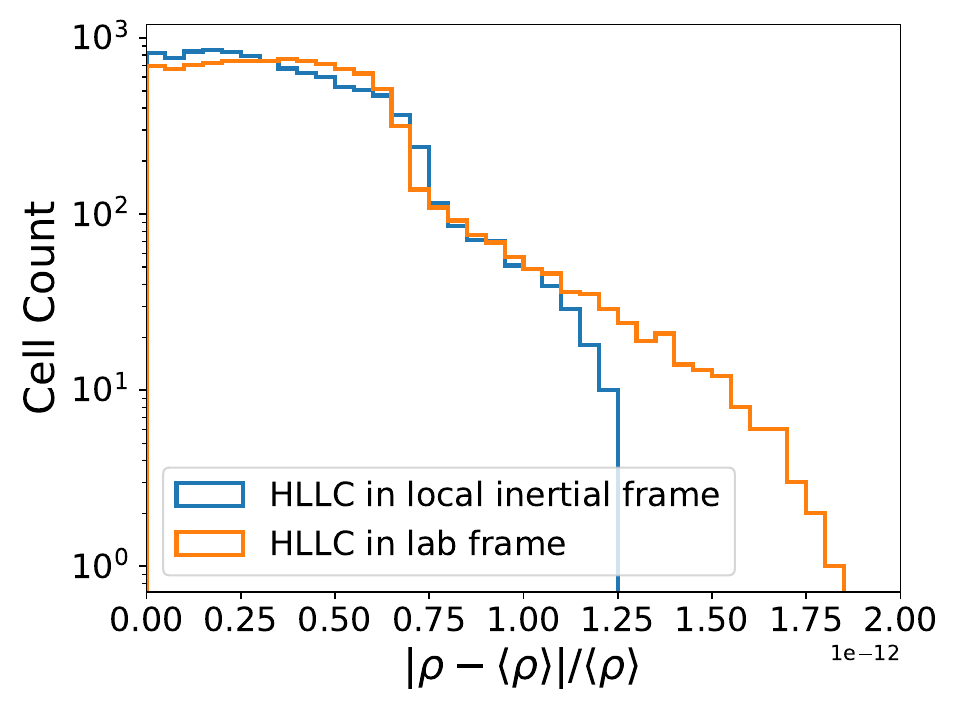}
    \caption{Histogram of density fluctuations at $t=t_{f}$ for the HLLC solver (orange) and the HLLC solver with the upwind-preserving scheme (blue).
    \label{fig:denshist}}
\end{figure}

We test the above formulation of the flux in Sprout by simulating the homologous expansion of a uniform medium. We initialize a 2-D domain of dimensions $N_{x}=N_{y}=128$ and size $L_{x}=L_{y}=1$ containing gas with $\rho_{0}=1$, $P_{0}=10^{-4}/\gamma$, and $\gamma=5/3$, so that $c_{s,0}=10^{-2}$. The center of expansion is chosen as the lower left corner of the domain, and the fluid velocity is homologous: $\vel=\rad/t_{0}$. The mesh expands at a constant rate $H=1/t_{0}$ over the duration of the run from $t_{0}=1$ to $t_{f}=10$. If the Riemann solver is not upwind-preserving, we would expect to see density fluctuations owing to sign errors in the mass flux. In Fig.~\ref{fig:denshist} we show the fractional density fluctuations about the mean for the HLLC solver when the flux is solved in the lab frame versus the local rest frame, ignoring fluid near the boundaries to eliminate any potential complications due to boundary conditions. We find that the fluctuations are larger by $\approx$40\% when solved in the lab frame.

\bibliography{references}{}

\begin{thebibliography}{}
\expandafter\ifx\csname natexlab\endcsname\relax\def\natexlab#1{#1}\fi
\providecommand{\url}[1]{\href{#1}{#1}}
\providecommand{\dodoi}[1]{doi:~\href{http://doi.org/#1}{\nolinkurl{#1}}}
\providecommand{\doeprint}[1]{\href{http://ascl.net/#1}{\nolinkurl{http://ascl.net/#1}}}
\providecommand{\doarXiv}[1]{\href{https://arxiv.org/abs/#1}{\nolinkurl{https://arxiv.org/abs/#1}}}

\bibitem[{{Arunachalam} {et~al.}(2022){Arunachalam}, {Hughes}, {Hovey}, \&
  {Eriksen}}]{2022ApJ...938..121A}
{Arunachalam}, P., {Hughes}, J.~P., {Hovey}, L., \& {Eriksen}, K. 2022, \apj,
  938, 121, \dodoi{10.3847/1538-4357/ac927c}

\bibitem[{{Bauer} {et~al.}(2019){Bauer}, {White}, \&
  {Bildsten}}]{2019ApJ...887...68B}
{Bauer}, E.~B., {White}, C.~J., \& {Bildsten}, L. 2019, \apj, 887, 68,
  \dodoi{10.3847/1538-4357/ab4ea4}

\bibitem[{{Boehner} {et~al.}(2017){Boehner}, {Plewa}, \&
  {Langer}}]{2017MNRAS.465.2060B}
{Boehner}, P., {Plewa}, T., \& {Langer}, N. 2017, \mnras, 465, 2060,
  \dodoi{10.1093/mnras/stw2737}

\bibitem[{Boerner {et~al.}(2023)Boerner, Deems, Furlani, Knuth, \&
  Towns}]{access}
Boerner, T.~J., Deems, S., Furlani, T.~R., Knuth, S.~L., \& Towns, J. 2023, in
  Practice and Experience in Advanced Research Computing, PEARC '23 (New York,
  NY, USA: Association for Computing Machinery), 173–176,
  \dodoi{10.1145/3569951.3597559}

\bibitem[{{Boos} {et~al.}(2024){Boos}, {Townsley}, \&
  {Shen}}]{2024arXiv240108011B}
{Boos}, S.~J., {Townsley}, D.~M., \& {Shen}, K.~J. 2024, arXiv e-prints,
  arXiv:2401.08011, \dodoi{10.48550/arXiv.2401.08011}

\bibitem[{{Duffell}(2016)}]{rt1d}
{Duffell}, P.~C. 2016, \apj, 821, 76, \dodoi{10.3847/0004-637X/821/2/76}

\bibitem[{{El-Badry} {et~al.}(2023){El-Badry}, {Shen}, {Chandra}, {Bauer},
  {Fuller}, {Strader}, {Chomiuk}, {Naidu}, {Caiazzo}, {Rodriguez}, {Nagarajan},
  {Yamaguchi}, {Vanderbosch}, {Roulston}, {G{\"a}nsicke}, {Han}, {Burdge},
  {Filippenko}, {Brink}, \& {Zheng}}]{2023OJAp....6E..28E}
{El-Badry}, K., {Shen}, K.~J., {Chandra}, V., {et~al.} 2023, The Open Journal
  of Astrophysics, 6, 28, \dodoi{10.21105/astro.2306.03914}

\bibitem[{{Ferrand} {et~al.}(2022){Ferrand}, {Tanikawa}, {Warren}, {Nagataki},
  {Safi-Harb}, \& {Decourchelle}}]{2022ApJ...930...92F}
{Ferrand}, G., {Tanikawa}, A., {Warren}, D.~C., {et~al.} 2022, \apj, 930, 92,
  \dodoi{10.3847/1538-4357/ac5c58}

\bibitem[{{Fleischmann} {et~al.}(2020){Fleischmann}, {Adami}, \&
  {Adams}}]{2020JCoPh.42309762F}
{Fleischmann}, N., {Adami}, S., \& {Adams}, N.~A. 2020, Journal of
  Computational Physics, 423, 109762, \dodoi{10.1016/j.jcp.2020.109762}

\bibitem[{{Garc{\'\i}a-Senz} {et~al.}(2012){Garc{\'\i}a-Senz}, {Badenes}, \&
  {Serichol}}]{2012ApJ...745...75G}
{Garc{\'\i}a-Senz}, D., {Badenes}, C., \& {Serichol}, N. 2012, \apj, 745, 75,
  \dodoi{10.1088/0004-637X/745/1/75}

\bibitem[{{Godinaud} {et~al.}(2023){Godinaud}, {Acero}, {Decourchelle}, \&
  {Ballet}}]{2023A&A...680A..80G}
{Godinaud}, L., {Acero}, F., {Decourchelle}, A., \& {Ballet}, J. 2023, \aap,
  680, A80, \dodoi{10.1051/0004-6361/202346954}

\bibitem[{{Gray} {et~al.}(2016){Gray}, {Raskin}, \&
  {Owen}}]{2016ApJ...833...62G}
{Gray}, W.~J., {Raskin}, C., \& {Owen}, J.~M. 2016, \apj, 833, 62,
  \dodoi{10.3847/1538-4357/833/1/62}

\bibitem[{Harris {et~al.}(2020)Harris, Millman, van~der Walt, Gommers,
  Virtanen, Cournapeau, Wieser, Taylor, Berg, Smith, Kern, Picus, Hoyer, van
  Kerkwijk, Brett, Haldane, del R{\'{i}}o, Wiebe, Peterson,
  G{\'{e}}rard-Marchant, Sheppard, Reddy, Weckesser, Abbasi, Gohlke, \&
  Oliphant}]{harris2020arrayNUMPY}
Harris, C.~R., Millman, K.~J., van~der Walt, S.~J., {et~al.} 2020, Nature, 585,
  357, \dodoi{10.1038/s41586-020-2649-2}

\bibitem[{{Hirai} {et~al.}(2018){Hirai}, {Podsiadlowski}, \&
  {Yamada}}]{2018ApJ...864..119H}
{Hirai}, R., {Podsiadlowski}, P., \& {Yamada}, S. 2018, \apj, 864, 119,
  \dodoi{10.3847/1538-4357/aad6a0}

\bibitem[{Hunter(2007)}]{Hunter:2007}
Hunter, J.~D. 2007, Computing in Science \& Engineering, 9, 90,
  \dodoi{10.1109/MCSE.2007.55}

\bibitem[{{Kasen}(2010)}]{2010ApJ...708.1025K}
{Kasen}, D. 2010, \apj, 708, 1025, \dodoi{10.1088/0004-637X/708/2/1025}

\bibitem[{{Liu} {et~al.}(2023){Liu}, {R{\"o}pke}, \&
  {Han}}]{2023RAA....23h2001L}
{Liu}, Z.-W., {R{\"o}pke}, F.~K., \& {Han}, Z. 2023, Research in Astronomy and
  Astrophysics, 23, 082001, \dodoi{10.1088/1674-4527/acd89e}

\bibitem[{{Liu} {et~al.}(2015){Liu}, {Tauris}, {R{\"o}pke}, {Moriya},
  {Kruckow}, {Stancliffe}, \& {Izzard}}]{2015A&A...584A..11L}
{Liu}, Z.-W., {Tauris}, T.~M., {R{\"o}pke}, F.~K., {et~al.} 2015, \aap, 584,
  A11, \dodoi{10.1051/0004-6361/201526757}

\bibitem[{{Lopez} {et~al.}(2011){Lopez}, {Ramirez-Ruiz}, {Huppenkothen},
  {Badenes}, \& {Pooley}}]{2011ApJ...732..114L}
{Lopez}, L.~A., {Ramirez-Ruiz}, E., {Huppenkothen}, D., {Badenes}, C., \&
  {Pooley}, D.~A. 2011, \apj, 732, 114, \dodoi{10.1088/0004-637X/732/2/114}

\bibitem[{{Mandal} \& {Duffell}(2023)}]{sprout}
{Mandal}, S., \& {Duffell}, P.~C. 2023, \apjs, 269, 30,
  \dodoi{10.3847/1538-4365/acfc19}

\bibitem[{{Mandal} {et~al.}(2023){Mandal}, {Duffell}, {Polin}, \&
  {Milisavljevic}}]{2023ApJ...956..130M}
{Mandal}, S., {Duffell}, P.~C., {Polin}, A., \& {Milisavljevic}, D. 2023, \apj,
  956, 130, \dodoi{10.3847/1538-4357/acf9fb}

\bibitem[{{Mandal} {et~al.}(2024){Mandal}, {Duffell}, {Polin}, \&
  {Milisavljevic}}]{2024arXiv240312264M}
---. 2024, arXiv e-prints, arXiv:2403.12264, \dodoi{10.48550/arXiv.2403.12264}

\bibitem[{{Marietta} {et~al.}(2000){Marietta}, {Burrows}, \&
  {Fryxell}}]{2000ApJS..128..615M}
{Marietta}, E., {Burrows}, A., \& {Fryxell}, B. 2000, \apjs, 128, 615,
  \dodoi{10.1086/313392}

\bibitem[{{Mignone}(2014)}]{2014JCoPh.270..784M}
{Mignone}, A. 2014, Journal of Computational Physics, 270, 784,
  \dodoi{10.1016/j.jcp.2014.04.001}

\bibitem[{{Mihalas} \& {Mihalas}(1984)}]{1984oup..book.....M}
{Mihalas}, D., \& {Mihalas}, B.~W. 1984, {Foundations of radiation
  hydrodynamics}

\bibitem[{{Ni} {et~al.}(2023){Ni}, {Moon}, {Drout}, {Matzner}, {Leong}, {Kim},
  {Park}, \& {Lee}}]{2023ApJ...959..132N}
{Ni}, Y.~Q., {Moon}, D.-S., {Drout}, M.~R., {et~al.} 2023, \apj, 959, 132,
  \dodoi{10.3847/1538-4357/ad0640}

\bibitem[{{Olling} {et~al.}(2015){Olling}, {Mushotzky}, {Shaya}, {Rest},
  {Garnavich}, {Tucker}, {Kasen}, {Margheim}, \&
  {Filippenko}}]{2015Natur.521..332O}
{Olling}, R.~P., {Mushotzky}, R., {Shaya}, E.~J., {et~al.} 2015, \nat, 521,
  332, \dodoi{10.1038/nature14455}

\bibitem[{{Papish} {et~al.}(2015){Papish}, {Soker}, {Garc{\'\i}a-Berro}, \&
  {Aznar-Sigu{\'a}n}}]{2015MNRAS.449..942P}
{Papish}, O., {Soker}, N., {Garc{\'\i}a-Berro}, E., \& {Aznar-Sigu{\'a}n}, G.
  2015, \mnras, 449, 942, \dodoi{10.1093/mnras/stv337}

\bibitem[{{Picquenot} {et~al.}(2024){Picquenot}, {Holland-Ashford}, \&
  {Williams}}]{2024A&A...687A..28P}
{Picquenot}, A., {Holland-Ashford}, T., \& {Williams}, B.~J. 2024, \aap, 687,
  A28, \dodoi{10.1051/0004-6361/202449155}

\bibitem[{{Seitenzahl} {et~al.}(2019){Seitenzahl}, {Ghavamian}, {Laming}, \&
  {Vogt}}]{2019PhRvL.123d1101S}
{Seitenzahl}, I.~R., {Ghavamian}, P., {Laming}, J.~M., \& {Vogt}, F.~P.~A.
  2019, \prl, 123, 041101, \dodoi{10.1103/PhysRevLett.123.041101}

\bibitem[{{Shen} {et~al.}(2018{\natexlab{a}}){Shen}, {Kasen}, {Miles}, \&
  {Townsley}}]{2018ApJ...854...52S}
{Shen}, K.~J., {Kasen}, D., {Miles}, B.~J., \& {Townsley}, D.~M.
  2018{\natexlab{a}}, \apj, 854, 52, \dodoi{10.3847/1538-4357/aaa8de}

\bibitem[{{Shen} {et~al.}(2018{\natexlab{b}}){Shen}, {Boubert}, {G{\"a}nsicke},
  {Jha}, {Andrews}, {Chomiuk}, {Foley}, {Fraser}, {Gromadzki}, {Guillochon},
  {Kotze}, {Maguire}, {Siebert}, {Smith}, {Strader}, {Badenes}, {Kerzendorf},
  {Koester}, {Kromer}, {Miles}, {Pakmor}, {Schwab}, {Toloza}, {Toonen},
  {Townsley}, \& {Williams}}]{2018ApJ...865...15S}
{Shen}, K.~J., {Boubert}, D., {G{\"a}nsicke}, B.~T., {et~al.}
  2018{\natexlab{b}}, \apj, 865, 15, \dodoi{10.3847/1538-4357/aad55b}

\bibitem[{{Shields} {et~al.}(2023){Shields}, {Arunachalam}, {Kerzendorf},
  {Hughes}, {Biriouk}, {Monk}, \& {Buchner}}]{2023ApJ...950L..10S}
{Shields}, J.~V., {Arunachalam}, P., {Kerzendorf}, W., {et~al.} 2023, \apjl,
  950, L10, \dodoi{10.3847/2041-8213/acd6a0}

\bibitem[{Stone {et~al.}(2020)Stone, Tomida, White, \& Felker}]{athena++}
Stone, J.~M., Tomida, K., White, C.~J., \& Felker, K.~G. 2020, The
  Astrophysical Journal Supplement Series, 249, 4,
  \dodoi{10.3847/1538-4365/ab929b}

\bibitem[{{Tanikawa} {et~al.}(2018){Tanikawa}, {Nomoto}, \&
  {Nakasato}}]{2018ApJ...868...90T}
{Tanikawa}, A., {Nomoto}, K., \& {Nakasato}, N. 2018, \apj, 868, 90,
  \dodoi{10.3847/1538-4357/aae9ee}

\bibitem[{{Tanikawa} {et~al.}(2019){Tanikawa}, {Nomoto}, {Nakasato}, \&
  {Maeda}}]{2019ApJ...885..103T}
{Tanikawa}, A., {Nomoto}, K., {Nakasato}, N., \& {Maeda}, K. 2019, \apj, 885,
  103, \dodoi{10.3847/1538-4357/ab46b6}

\bibitem[{{Taylor}(1950)}]{1950RSPSA.201..159T}
{Taylor}, G. 1950, Proceedings of the Royal Society of London Series A, 201,
  159, \dodoi{10.1098/rspa.1950.0049}

\bibitem[{{Teyssier}(2002)}]{ramses}
{Teyssier}, R. 2002, \aap, 385, 337, \dodoi{10.1051/0004-6361:20011817}

\bibitem[{{Toro} {et~al.}(1994){Toro}, {Spruce}, \&
  {Speares}}]{1994ShWav...4...25T}
{Toro}, E.~F., {Spruce}, M., \& {Speares}, W. 1994, Shock Waves, 4, 25,
  \dodoi{10.1007/BF01414629}

\bibitem[{{Truelove} \& {McKee}(1999)}]{1999ApJS..120..299T}
{Truelove}, J.~K., \& {McKee}, C.~F. 1999, \apjs, 120, 299,
  \dodoi{10.1086/313176}

\bibitem[{{Vink}(2012)}]{2012A&ARv..20...49V}
{Vink}, J. 2012, \aapr, 20, 49, \dodoi{10.1007/s00159-011-0049-1}

\bibitem[{Virtanen {et~al.}(2020)Virtanen, Gommers, Oliphant, Haberland, Reddy,
  Cournapeau, Burovski, Peterson, Weckesser, Bright, {van der Walt}, Brett,
  Wilson, Millman, Mayorov, Nelson, Jones, Kern, Larson, Carey, Polat, Feng,
  Moore, {VanderPlas}, Laxalde, Perktold, Cimrman, Henriksen, Quintero, Harris,
  Archibald, Ribeiro, Pedregosa, {van Mulbregt}, \& {SciPy 1.0
  Contributors}}]{2020SciPy-NMeth}
Virtanen, P., Gommers, R., Oliphant, T.~E., {et~al.} 2020, Nature Methods, 17,
  261, \dodoi{10.1038/s41592-019-0686-2}

\bibitem[{{Williams} {et~al.}(2016){Williams}, {Chomiuk}, {Hewitt}, {Blondin},
  {Borkowski}, {Ghavamian}, {Petre}, \& {Reynolds}}]{2016ApJ...823L..32W}
{Williams}, B.~J., {Chomiuk}, L., {Hewitt}, J.~W., {et~al.} 2016, \apjl, 823,
  L32, \dodoi{10.3847/2041-8205/823/2/L32}

\bibitem[{{Wong} {et~al.}(2024){Wong}, {White}, \&
  {Bildsten}}]{2024arXiv240800125W}
{Wong}, T. L.~S., {White}, C., \& {Bildsten}, L. 2024, arXiv e-prints,
  arXiv:2408.00125.
\newblock \doarXiv{2408.00125}

\bibitem[{{XRISM Collaboration}(2024)}]{2024arXiv240814301X}
{XRISM Collaboration}. 2024, arXiv e-prints, arXiv:2408.14301,
  \dodoi{10.48550/arXiv.2408.14301}

\end{thebibliography}
\bibliographystyle{aasjournal}


\end{document}